\documentclass[3p,times]{elsarticle}
\usepackage{graphicx}
\usepackage{sidecap}
\usepackage{amssymb,amsmath,amsthm,amsbsy}
\usepackage{calligra,calrsfs,mathrsfs}
\usepackage{multirow}
\usepackage{ulem,cancel}
\usepackage[usenames,dvipsnames]{color}
\usepackage{copyrightbox}
\usepackage{cases}
\usepackage{pgfplots}
\usepackage{subcaption}
\usepackage{tikz} 
\usepackage{tkz-euclide}
\usepackage{tikz-dimline}
\usetikzlibrary[patterns]
\usepackage{hyperref}
\usepackage{algorithm}
\usepackage{algpseudocode}
\usepackage{dsfont}
\usepackage{caption}
\usepackage{epstopdf}
\usepackage{epsfig}
\usepackage{natbib}
\usepackage{grffile}
\usepackage{comment}
\usepackage{enumerate}
\usepackage{array}
\newcolumntype{P}[1]{>{\centering\arraybackslash}p{#1}}
\usepackage{float}
\usepackage[T1]{fontenc}
\usepackage[latin9]{inputenc}
\usepackage{geometry}
\usepgfplotslibrary{fillbetween}
\usetikzlibrary{patterns}
\usepackage{color}
\usepackage{stmaryrd}
\usepackage{setspace}
\usepackage{amsthm}
\usepackage{textcomp}
\usepackage{hyperref}%
\usepackage{soul,xcolor}
\usepackage[inline]{enumitem}
\usepackage{textcomp}
\usepackage{adjustbox}
\usepackage{todonotes}
\usepackage{empheq}
\graphicspath{{./figs/}}

\allowdisplaybreaks

\newcommand{\ifcomment}{\iffalse}

\newcommand{\dov}[1]{\frac{d}{d{#1}}}

\newcommand{\dd}[2]{\frac{d{#1}}{d{#2}}}

\newcommand{\avg  }[1]{\langle #1 \rangle}

\newdefinition{rem}{Remark}

\biboptions{sort&compress}

\journal{Not Yet}

\begin{document}

\begin{frontmatter}

\title{The Ganglion Network Model:\\ Evolving Trapped Phases in Porous Media}
\author[psu]{Yashar Mehmani\corref{ca}}
\ead{yzm5192@psu.edu}
\cortext[ca]{Corresponding author}
\address[psu]{Department of Energy and Mineral Engineering, The Pennsylvania State University, University Park, USA}

\begin{abstract}
Partially miscible ganglia trapped within porous media, spanning one or multiple pores and evolving through diffusive mass transfer, are common in subsurface (e.g., CO$_2$ and H$_2$ storage) and manufacturing (e.g., fuel cells) applications.
We present the \textit{ganglion network model} (GNM), a reduced-order method for simulating how a population of such ganglia evolves inside an arbitrary porous microstructure.
GNM operates on a tree graph, the \textit{ganglion network}, extracted from the pore-scale image of a given sample. Each point on the graph encodes a possible ganglion configuration in the void space, without any loss of geometric or topological complexity. The evolution of a population is modeled by representing each ganglion as a particle on the graph and tracking it according to a set of rules formulated herein. The rules capture capillary events such as pore invasion, retraction, snap-off, fragmentation, and merger. Unlike pore-network models, another graph-based modeling tool at the pore scale, GNM solves no system of equations and its cost scales with ganglion count, not domain size.
We validate GNM against an image-based pore-network model in 2.5D and 3D domains with populations undergoing ripening, dissolution, and growth. We find good agreement in ganglion statistics, aggregate properties, and spatial configuration.
We further argue that the ganglion network is the statistical space needed for extending kinetic theories of Ostwald ripening from single- to multi-pore ganglia, and provide an outline for how to do this.
GNM opens the door to modeling other dynamics of trapped phases in porous media.
\end{abstract}


\end{frontmatter}

\section{Introduction}
\label{sec:intro}

A \textit{ganglion} is a disconnected blob of non-wetting fluid trapped inside a porous medium otherwise filled with a wetting fluid. Ganglia are ubiquitous in subsurface and manufacturing applications. In underground hydrogen storage, repeated injection--withdrawal cycles to/from an aquifer can cause H$_2$ to be trapped as ganglia~\cite{zivar2021uhs}. In geologic carbon sequestration, such entrapment is desired because it immobilizes the CO$_2$~\cite{bachu2008co2}. Ganglia also arise in petroleum reservoirs, where pressure decline near wells causes gas bubbles to exsolve from oil and thwart production~\cite{gao2021bubble}. In groundwater, ganglia are relevant in the remediation of residually trapped non-aqueous phase liquids (NAPL)~\cite{imhoff1996NAPL} and the oxygenation of shallow aquifers by trapped air bubbles post-rainfall~\cite{holocher2003dissol}. In fuel cells and electrolyzers, porous components must be continuously purged of internally generated bubbles and droplets that degrade performance~\cite{andersson2016FC, lee2020FCbazyl}. What ties these applications together is that ganglia are partially miscible in the wetting fluid and exchange mass by molecular diffusion. As they shrink or grow, ganglia may invade new pores, retract from occupied ones, snap off at throats, fragment, or merge. Here, we seek to predict how such ganglia evolve in arbitrary porous microstructures.

The solute concentration in equilibrium with a ganglion is set by its interfacial curvature, $\kappa$. The Young--Laplace equation ties $\kappa$ to the ganglion's internal pressure, and an equation of state (here Henry's law) ties that to concentration.
The higher the $\kappa$, the more elevated the interface-adjacent concentration is. If the wetting fluid is saturated, mass released by high-$\kappa$ ganglia is fully absorbed by low-$\kappa$ ones, a phenomenon called \textit{Ostwald ripening}~\cite{ostwald1897} in which the wetting fluid acts merely as a transfer medium without storage. Were the wetting fluid undersaturated or supersaturated, ganglia would, on average, shrink or grow, respectively, while still exchanging mass. We refer to these scenarios as \textit{dissolution} and \textit{growth}, in which the wetting fluid also stores solute. Despite lacking storage, ripening is more challenging to model, because it is driven solely by capillary-pressure differences between ganglia, with no net exchange with the wetting fluid to mask errors. The understanding of ripening in bulk fluids is mature, since bubbles are spherical with $\kappa$ inversely proportional to radius, causing any population to coarsen into a single bubble at equilibrium~\cite{voorhees1985, bray2002theory}. In porous media, research is nascent and ripening is more complex, because confinement deforms ganglia and controls $\kappa$. Even for a single-pore occupying bubble, the curvature--volume relation $\kappa(V_b)$ is U-shaped~\cite{xu2017PRL}, leading populations to equilibrate at a common $\kappa$ without coarsening into one bubble~\cite{xu2017PRL, deChalendar2018pnm, mehmani2022JCP}. For multi-pore ganglia, which are far more common in nature~\cite{iglauer2011gang, garing2017PcExp}, $\kappa(V_b)$ is further complicated by oscillations and discontinuities~\cite{wang2021PNAS, mehmani2022AWR} caused by capillary events that can abruptly alter a ganglion's topology, pore occupancy, and interfacial geometry.\looseness=-1

Ripening in porous media has been probed with microfluidic experiments~\cite{xu2017PRL, joewondo2023exp, salehpour2025micro}, X-ray micro-CT imaging of rocks~\cite{zhang2023trap, goodarzi2024exp, boon2024XrayH2, song2025XrayH2}, pore-network models (PNM)~\cite{deChalendar2018pnm, mehmani2022JCP, mehmani2022AWR, bueno2023PNM, laku2026ipnm, adebimpe2026pnm}, and level-set simulations~\cite{singh2022level, singh2023pore, singh2024ostwald}.
Based on these, continuum theories have been developed to evolve ganglion saturation~\cite{xu2019GravGRL, yaxin2020JFM, yaxin2022TIPM, feng2022KeGravTIPM, blunt2022ostwald, mehmani2024deplete}, equilibrium theories to predict final ganglion statistics from initial ones~\cite{mehmani2022JCP, bueno2023PNM}, and kinetic theories to predict how such statistics evolve over time~\cite{yu2023GRLkinetics, bueno2024theory, bueno2025theory}.
Dissolution and growth of (e.g., NAPL, H$_2$, CO$_2$) ganglia have been studied with similar tools~\cite{sahloul2002NAPL, chomsurin2003NAPL, jangda2023pore, yortsos1999visual, dillard2000NAPL, held2001pnm, joewondo2022pnm, yortsos1995visual, dominguez2000gas, ioannidis2011gas, berg2020PressDeplete}. However, the PNMs used tend to simplify or neglect features of the $\kappa(V_b)$ inside pores that are critical to ripening.
Early ripening PNMs were limited to single-pore bubbles~\cite{deChalendar2018pnm, mehmani2022JCP}. The first extension to multi-pore ganglia occurred in~\cite{mehmani2022AWR}, but assumed a stagnant wetting phase, a single simplified pore shape, and instantaneous capillary equilibrium of each ganglion (i.e., no internal flow).
All three limitations were lifted by the image-based PNM (iPNM) of Laku et al.~\cite{laku2026ipnm}, where each pore's $\kappa(V_b)$ is derived from its local sub-image using a modified pore-morphology method (PMM)~\cite{hazlett1995pmm, hilpert2001pmm, mehmani2024deplete} encoding the pore's actual shape. Moreover, internal flow within each phase is computed and, depending on which part of the ganglion a pore represents, its $\kappa(V_b)$ is adapted dynamically during simulations. iPNM was validated against recent microfluidic experiments of H$_2$ ripening~\cite{salehpour2025micro}, and we use it here as a reference.

Pore-network models aim to compute the state of each pore in physical space (e.g., pressure, saturation), whereas kinetic theories compute the state of each ganglion in a statistical \textit{phase space}. The present work builds on this latter conceptualization, which we examine next.
In the classical Lifshitz--Slyozov--Wagner (LSW) theory for bulk fluids~\cite{lifshitz1961orig, wagner1961orig}, a bubble's state is determined by its volume $V_b$ (or radius), and the phase space consists of the positive real line. Hence, the number density of bubble states, $f(V_b,t)$, evolves according to the population balance:
\begin{equation}\label{eq:pbe}
\frac{\partial f}{\partial t} + \frac{\partial}{\partial V_b}\left(f\,\frac{dV_b}{dt}\right) = 0,
\end{equation}
where the \textit{phase velocity} is given by:
\begin{equation}\label{eq:lswclosure}
\frac{dV_b}{dt} \,\propto\, R_b\,(\kappa_m - \kappa_b).
\end{equation}
The $R_b$ and $\kappa_b$ denote the radius and curvature of a bubble, and $\kappa_m$ is the mean-field (or critical) curvature in equilibrium with the far-field concentration. Bubbles with $\kappa_b \!>\! \kappa_m$ shrink, and the rest grow.
For spherical bubbles inside a homogeneous porous medium, Yu et al.~\cite{yu2023GRLkinetics} adapted Eq.~\ref{eq:lswclosure} by replacing $R_b$ with $A_t/L_c$, the throat cross-sectional area over a mass-transfer length.
Bueno et al.~\cite{bueno2024theory} extended this to deformed bubbles in heterogeneous, spatially correlated porous media, and showed the phase space now consists of two axes, pore size $R_p$ and $V_b$.
A later extension to $n_c$-component bubbles found that the phase space consists of $n_c+1$ axes~\cite{bueno2025theory}. All existing kinetic theories assume that bubbles occupy a single pore and that the wetting fluid stores no solute, limiting them to modeling ripening only.

A general kinetic theory for multi-pore ganglia remains an open problem, and the obstacle is the phase space itself.
We claim it cannot be Euclidean as in existing theories. To see this, consider what it takes to uniquely determine the state of a ganglion occupying $N_p$ pores.
Aside from $V_b$, we would need $N_p$ extra axes, one for each occupied pore size. We also need another $\tbinom{N_p}{2}$ axes, one for each possible throat size connecting the pores. Since only a subset of throats may be invaded, it is convenient to adopt the convention that \textit{zero} on a throat axis means \textit{vacant}.
A ganglion positioned at any point of this space has a unique $\kappa_b$, obtained by imposing mass conservation and uniform curvature~\cite{mehmani2022AWR}.
The trouble is that for a 100-pore ganglion, this space has over 5000 axes, and it is not even the entire phase space. As the ganglion grows or shrinks, according to an equation like Eq.~\ref{eq:lswclosure}, its pore occupancy can change due to capillary events. For example, the ganglion may fragment into $n$ pieces, each with a smaller occupancy $N_{p,i}$, $i\!\in\!\{1,\cdots,n\}$. Those pieces now live in their own separate spaces of dimension ${1+N_{p,i}+\tbinom{N_{p,i}}{2}}$. Therefore, the phase space has an axis for $N_p$, consisting of the positive integers, at each fixed value of which the corresponding subspace has $\smash{1+N_{p}+\tbinom{N_{p}}{2}}$ axes. The sheer number of subspace dimensions, to say nothing of how ganglion states are to move between them, is sufficient to abandon such an extension, a limitation characterized in~\cite{bueno2025theory} as \textit{foundational} to existing theories.

We posit that the most parsimonious phase space for multi-pore ganglia in arbitrary porous microstructures is a tree graph. We extract this graph directly from a pore-scale image and call it the \textit{ganglion network}. Every point on the graph is a possible ganglion state, whose volume, curvature, and occupied void space---hence pore occupancy, topology, and interfacial geometry---are fully determined.
Conversely, every possible ganglion that fits within the void space is represented by a point on the graph.
We construct the ganglion network through an adaptation of PMM, where the void space is morphologically opened with structuring elements of successively increasing radii. The connected components that survive each radius $r$ are \textit{all} ganglia with curvature $\kappa_b\!=\!2/r$. By tracking these connected components, and recording which ganglia descend from the fragmentation of a parent ganglion at smaller $r$, the tree is built.
Snap-off and entry curvatures of the occupied throats of each ganglion are encoded at the junctions, and extra links are appended at leaf nodes to capture small, spherical ganglia.
By storing the image pixels corresponding to each ganglion at the graph's nodes, any ganglion state on the graph is mappable onto the pore-scale image. This mapping allows the spatial distribution of a population, represented by a set of points on the graph, to be visualized.

The ganglion network admits two uses: (1) as a kinetic theory in the tradition of LSW for evolving the statistics of a population; and (2) as a reduced-order model for evolving ganglia in a specific porous sample. We pursue the latter in this work and present the \textit{ganglion network model} (GNM), then discuss the former in Section~\ref{sec:disc_theory}. In GNM, each ganglion is represented by a point on the ganglion network, and the wetting fluid is modeled as a mean field with solute concentration $X_m$. The mean field can store solute, enabling GNM to model ripening, dissolution, and growth. Each ganglion moves along the link it occupies on the tree graph at phase velocity:
\begin{equation}\label{eq:gnmclosure}
\frac{dV_b}{dt} \,\propto\, \frac{\langle A\rangle}{\langle L\rangle}\,(\kappa_m - \kappa_b),
\end{equation}
which captures the mass exchange between the ganglion and the mean field. Eq.~\ref{eq:gnmclosure} has the same form as Eq.~\ref{eq:lswclosure}, with $R_b$ replaced by the average throat area over the mean ganglion spacing, $\langle A\rangle/\langle L\rangle$. The $\kappa_m$ is derived from $X_m$, evolved in turn by a separate balance equation.
Upon reaching a junction in the graph, a shrinking ganglion fragments into the branches below, whereas a growing ganglion spills into neighboring branches---mimicking snap-off and pore invasion.
Evolving a population thus reduces to moving a set of points on the ganglion network according to rules we formulate herein.
Similar to PNM, which is a graph-based, reduced-order abstraction of the pore space, GNM is a graph-based, reduced-order abstraction of ganglion states---trading pore-centric calculations for ganglion-centric ones.
But unlike PNM, no system of equations is solved in GNM and the cost scales with the number of ganglia, not domain size.

We validate GNM against iPNM~\cite{laku2026ipnm}, which resolves the concentration field and the evolution of ganglia explicitly, to test the mean-field approximation and the rules of ganglion motion on the tree graph.
We consider 2.5D (planar with out-of-plane thickness) and 3D domains with complex microstructures, each subjected to the ripening, dissolution, or growth of an initial population.
The version of iPNM used here improves upon~\cite{laku2026ipnm} by making its solute balance equation mass conservative (\ref{app:ipnm}).
Because iPNM, like GNM, maps ganglia onto the pore-scale image without geometric simplification, predictions of the spatial configuration of ganglia can be compared one to one.

The paper's outline follows. Section~\ref{sec:probdesc} describes the problem. Section~\ref{sec:gnm} formulates GNM, including the extraction of the ganglion network (Section~\ref{sec:extract}), the mean-field equations (Section~\ref{sec:meanfield}), and the rules of ganglion motion on the graph (Section~\ref{sec:rules}), followed by details on the numerical solution and visualization. Section~\ref{sec:valid} presents the domains and scenarios used to validate GNM against iPNM, and Section~\ref{sec:results} compares predictions from both. Section~\ref{sec:disc} discusses sources of discrepancy, computational complexity, limitations, and future extensions. Section~\ref{sec:conc} concludes the paper.

\section{Problem description}
\label{sec:probdesc}

\begin{figure}[t!]
\centering
\includegraphics[width=\linewidth,trim={0 0.3cm 0 0},clip]{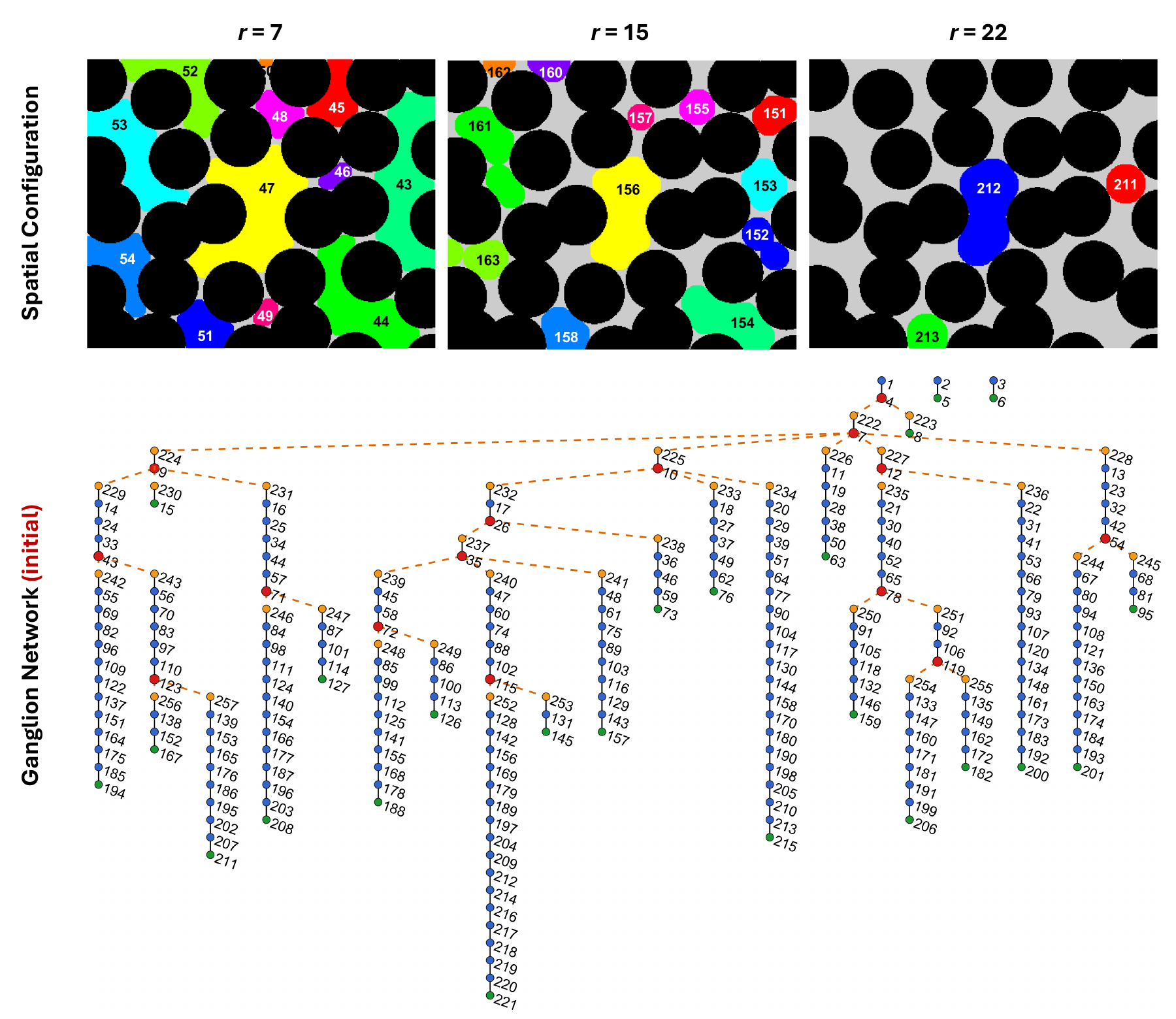}
\caption{Construction of the ganglion network from a binary image of a 2D disc pack (solid: black, void: gray). Top row: the image opened with structuring elements of radius $r=7$, $15$, and $22$ pixels. Each connected component of the opened image is a possible ganglion configuration and is colored and numbered by its node in the network. Bottom row: the resulting ganglion network, a tree graph whose nodes are regular (blue), junction (red), virtual (orange), and leaf (green). Solid lines are regular links and dashed lines are virtual links. Terminal nodes are not drawn. The network corresponds to the \textit{initial graph} constructed in Section~\ref{sec:extract} before adjustment to junction/virtual nodes, addition of leaf links, and coarsening.}
\label{fig:construct}
\end{figure}

We consider a rigid porous medium whose void space is occupied by a wetting liquid (water) and a population of trapped non-wetting ganglia, as illustrated in one of the top panels of Fig.~\ref{fig:construct}. The geometry of the solid is characterized by a binary image (solid: black, void: gray), and ganglia vary in size, spanning one or more pores. The ganglia consist of a single species of constant, uniform density that is partially miscible in the water, which can store it as dissolved solute. No flow is externally imposed, the system is isothermal, and the contact angle between every ganglion and the solid is zero. Depending on whether the water is supersaturated, undersaturated, or saturated with the solute, the population can collectively grow, dissolve, or undergo Ostwald ripening. All three scenarios are considered herein.

Mass exchange between the ganglia and the water is driven by the interfacial curvature of the ganglia. We assume local thermodynamic equilibrium at a ganglion's interface, where the dissolved concentration obeys Henry's law:
\begin{equation}
X_b = \frac{p_b - p_v}{H}.
\label{eq:henry}
\end{equation}
Here, $X_b$ is the solute mole fraction adjacent to the ganglion, $p_b$ is the ganglion's pressure, $p_v$ is the partial pressure of water that may have vaporized into the ganglion, and $H$ is Henry's constant. Interfacial tension causes $p_b$ to become elevated compared to the ambient water pressure $p_w$, as dictated by the Young--Laplace equation:
\begin{equation}
p_b = p_w + \sigma \kappa_b,
\label{eq:younglaplace}
\end{equation}
where $\sigma$ is the surface tension and $\kappa_b$ is the curvature of the ganglion's interface. Combining Eqs.~\ref{eq:henry}--\ref{eq:younglaplace}, we see that a ganglion's $\kappa_b$ dictates the $X_b$ adjacent to it.
If water is undersaturated relative to all $X_b$, every ganglion dissolves, and if supersaturated, every ganglion grows.
At near saturation, differences in $\kappa_b$ among ganglia drive diffusive mass transfer from high to low curvature, with water acting merely as a transmission medium.
Since solute is assumed to be dilute, diffusion obeys Fick's law. We further assume no new ganglia nucleate and that the population evolves only through diffusion-induced growth/shrinkage and capillary events (e.g., pore invasion, snap-off, fragmentation, and merger).

\section{The ganglion network model (GNM)}
\label{sec:gnm}

\subsection{Extracting the ganglion network}
\label{sec:extract}

The ganglion network is constructed from a binary image via a modified pore-morphology method (PMM)~\cite{mehmani2024deplete,salehpour2025micro}. Classical PMM computes the capillary pressure--saturation relationship of a porous medium under drainage, and it discards any non-wetting phase disconnected from the inlet. The modified PMM retains the disconnected regions, which constitute possible configurations of a ganglion in the absence of external flow, consistent with the assumptions of Section~\ref{sec:probdesc}. In the following, we build on the modified PMM and use it to extract a tree graph corresponding to all possible ganglion configurations within the porous medium. We refer to this graph as the \textit{ganglion network}.
To simplify presentation, we first describe how an initial graph is extracted, followed by adjustments made to it afterwards.
\\

\noindent\textbf{Initial graph.}
Let $I$ denote the binary image, such as the one in Fig.~\ref{fig:construct}, with solid and void pixels valued 0 and 1, respectively. Let $B_r$ denote a disc-shaped structuring element in 2D, or a spherical one in 3D, of radius $r$. Morphological \textit{erosion} $I \ominus B_r$ is defined as the operation that retains every void pixel at which $B_r$, centered on that pixel, fits entirely within the void. Morphological \textit{dilation} $I \oplus B_r$ is the operation that grows the void pixels outward by a distance $r$. Morphological \textit{opening}, which is key to PMM, is thus defined by an erosion followed by a dilation:
\begin{equation}
O_r(I) = \left(I \ominus B_r\right) \oplus B_r.
\label{eq:opening}
\end{equation}
The resulting image is the portion of the void space accessible to the non-wetting phase whose interfacial radius of curvature is $r$. In 2.5D domains, $r$ is the in-plane component of the curvature. Consistent with Section~\ref{sec:probdesc}, the PMM employed here assumes a zero contact angle. Extensions to finite contact angles exist and are also applicable~\cite{schulz2015pmmCA,liu2022pmmCA}.

We are now well-positioned to describe the steps needed to extract the ganglion network, schematized by Fig.~\ref{fig:construct}.
To start, the distance map $\mathcal{D}$ of the void space is computed, which assigns to each void pixel its distance to the nearest solid surface. The maximum of $\mathcal{D}$, denoted $r_{max}$, is the largest inscribed radius that fits within the void space.
The tree graph of the ganglion network begins with root nodes that correspond to connected components of the void space in $I$, detected using a connected-component labeling algorithm. These components are then successively opened with $r=1$ to $r_{max}$ to construct the rest of the tree graph.
Concretely, let $O_0(I) = I$, representing opening with $r=0$, and proceed by induction.
For each $k\!\geq\!1$, open $O_{k-1}(I)$ with $B_{r_k}$ and detect the connected components of the resulting $O_k(I)$.
Each component is a possible configuration of a ganglion with radius of curvature $r_k$, and it is captured by a new node in the graph. These new nodes are the children of the nodes that represent connected components of $O_{k-1}(I)$. In other words, the ganglia in $O_k(I)$ are fathered by shrinking those in $O_{k-1}(I)$, hence $O_k(I)\subset O_{k-1}(I)$.

The shrinkage has three possible outcomes for a ganglion in $O_{k-1}(I)$ that determine how new nodes for $O_k(I)$ are created: (1) If the ganglion remains a single connected component after shrinking, the new node is the lone child of its parent and called a \textit{regular node}. The link connecting the two nodes is called a \textit{regular link}; (2) If the ganglion vanishes, no new nodes or links are created, and the node of the vanished ganglion is dubbed a \textit{leaf node}. A ganglion at a leaf node occupies only one pore and coincides with that pore's maximum inscribed disc or sphere;
(3) If the ganglion breaks up into $m$ fragments, the ganglion's node is dubbed a \textit{junction node} and $m$ new child nodes are created under it, one for each fragment. These child nodes are called \textit{virtual nodes} and the links connecting them to the junction node are called \textit{virtual links}. Underneath each virtual node we then create a regular node connected by a regular link. The regular nodes are the true nodes that represent each fragment, not the virtual nodes. Virtual nodes serve to ensure volume conservation and curvature continuity in the graph. Namely, the sum of ganglion volumes at virtual nodes equals the volume at the junction node, and virtual-node curvatures equal the junction-node curvature.

Each node $i$ in the tree stores three pieces of information about the ganglion it represents, captured by the triplet $(\kappa_i, V_i, \mathbf{x}_i)$. The first is the ganglion's curvature $\kappa_i$, obtained from the opening radius $r$ used to create it:
\begin{equation}
\kappa_i = \frac{1}{r} + \frac{2}{g} \quad (2.5\mathrm{D}),
\qquad
\kappa_i = \frac{2}{r} \quad (3\mathrm{D}),
\label{eq:nodecrv}
\end{equation}
where $g$ is the out-of-plane gap thickness. Second is the ganglion's volume $V_i$, obtained by summing the pixel volumes that comprise it. The third is the ganglion's position $\mathbf{x}_i$, computed as the centroid of those pixels. Moving from the top of the tree downward, both $\kappa_i$ and $V_i$ decrease, as ganglia shrink in size and their interfaces become less curved.
\\

\noindent\textbf{Leaf links.}
Recall a ganglion sitting at a leaf node coincides with the maximum inscribed disc or sphere of the sole occupied pore.
Following literature terminology~\cite{xu2017PRL, mehmani2022JCP}, we call such a ganglion \textit{critical}. If any larger, the ganglion gets deformed by the pore's walls and is called \textit{supercritical}. Any smaller, the ganglion remains disc-shaped or spherical as it detaches from the confining pore's walls and is called \textit{subcritical}.
Notice all nodes in the initial graph created above are either critical (leaf nodes) or supercritical (other nodes). To capture subcritical bubbles we append a new node underneath each leaf node, termed a \textit{terminal node}, with the following properties: $\kappa\!=\!\infty$, $V_i\!=\!0$, and $\mathbf{x}_i$ equaling the leaf node's position. Terminal nodes represent leaf-node ganglia at their moment of vanishing. The link connecting a leaf node to a terminal node is called a \textit{leaf link}, along which curvature depends on volume as follows:
\begin{equation}
\kappa = \sqrt{\frac{\pi g}{V}} + \frac{2}{g} \quad (2.5\mathrm{D}),
\qquad
\kappa = 2\left(\frac{4\pi}{3V}\right)^{1/3} \quad (3\mathrm{D}).
\label{eq:leafcrv}
\end{equation}
On all other links, the dependence follows the linear interpolation:
\begin{equation}
\kappa = \frac{\kappa_j - \kappa_i}{V_j - V_i} (V - V_i) + \kappa_i
\label{eq:intrpcrv}
\end{equation}
between the curvatures and volumes of nodes $i$ and $j$ that straddle the link in question. Notice leaf-links are the only links in the tree along which $\kappa$ grows monotonically as $V$ decreases, going down the graph. For all other links, the relation is monotonically decreasing.
Moreover, Eq.~\ref{eq:intrpcrv} yields $\kappa\!=\!\kappa_i\!=\!\kappa_j$ for virtual links. In GNM, a ganglion can reside on any node or link in the tree graph except on a virtual link, which is clarified further in Section~\ref{sec:rules}.
\\

\noindent\textbf{Junction nodes.}
In the initial graph extracted above, the curvature stored at each junction node $\kappa_j$ equals that of its virtual-node children $\kappa_v$, and the volume stored at the junction node $V_j$ equals the sum of the virtual-node volumes $V_v$. Since $\kappa_j$ is the curvature at which a shrinking ganglion fragments, it must correspond to the snap-off curvature $\kappa_s$ of the throat(s) that get(s) severed during fragmentation. However, this does not hold for the initial graph because $\kappa_j\!\neq\!\kappa_s$. We thus proceed to overwrite $\kappa_j$ with $\kappa_s$ so junction nodes represent ganglia at their true snap-off state.
In 3D, where a spherical structuring element is used to extract the initial graph, $\kappa_s\!=\!\kappa_j/2$ assuming throats are prisms.
In 2.5D, where opening is applied to a 2D image first followed by adding the out-of-plane curvature $2/g$ to nodes via Eq.~\ref{eq:nodecrv}:
\begin{equation}
\kappa_s = \max\left(\frac{1}{r_j},\ \frac{2}{g}\right) \quad (2.5\mathrm{D}),
\label{eq:snapcrv}
\end{equation}
where $r_j$ is the opening radius that created the junction node~\cite{lenormand1983Pcs}.
As a result, $\kappa_s\!<\!\kappa_j$ in both 2.5D and 3D.
After overwriting $\kappa_j$ with $\kappa_s$, we overwrite the junction-node volume $V_j$ with the value obtained by applying Eq.~\ref{eq:intrpcrv} to the parent link of the junction node at $\kappa\!=\!\kappa_s$. These overwrites effectively extend the parent link downward in the tree.
\\

\noindent\textbf{Virtual nodes.}
The curvature at a virtual node $\kappa_v$ of the initial graph equals the junction-node curvature $\kappa_j$, and the virtual-node volume $V_v$ is a fraction of the junction-node volume $V_j$. However, just like each junction node represents a ganglion prior to snap off, each virtual node represents a ganglion prior to invading its adjacent pore(s). If all virtual nodes of a junction contain a ganglion fragment, any increase in their volume will cause them to coalesce into a larger ganglion sitting at the junction node.
Hence, $\kappa_v$ must equal the capillary-entry curvature $\kappa_e$ of the connecting throat(s) between the fragments.
In 2.5D, this already holds for the initial graph built, because the additive expression for $\kappa_v$ ($=\!\kappa_j$) in Eq.~\ref{eq:nodecrv} agrees to high accuracy ($<$6\% error; see~\ref{app:entry}) with $\kappa_e$ of a prismatic throat. Hence, no overwrite of $\kappa_v$ with $\kappa_e$ is needed in 2.5D.
However in 3D, $\kappa_v\!=\!2\kappa_s$ ($=\!\kappa_j$) at a junction node, as argued in the ``Junction-node'' subsection above.
Computing $\kappa_e$ requires detecting throat(s) that connect ganglion fragments under each junction from the pore-scale image, then performing a non-trivial local analysis, e.g., Mayer--Stowe--Princen method~\cite{mayer1965MSP,princen1969MSP}.
Here, we forgo such an analysis in favor of a 
simpler heuristic: set $\kappa_e = 1.88 \kappa_s$, where the factor 1.88 holds for a square-shaped throat cross section at zero contact angle. We next proceed to overwrite $\kappa_v$ with $\kappa_e$, and overwrite the associated volume $V_v$ with the value obtained by applying Eq.~\ref{eq:intrpcrv} to the child link of the virtual node at $\kappa\!=\!\kappa_e$.
\\

\begin{figure}[t!]
\centering
\includegraphics[width=1.0\linewidth]{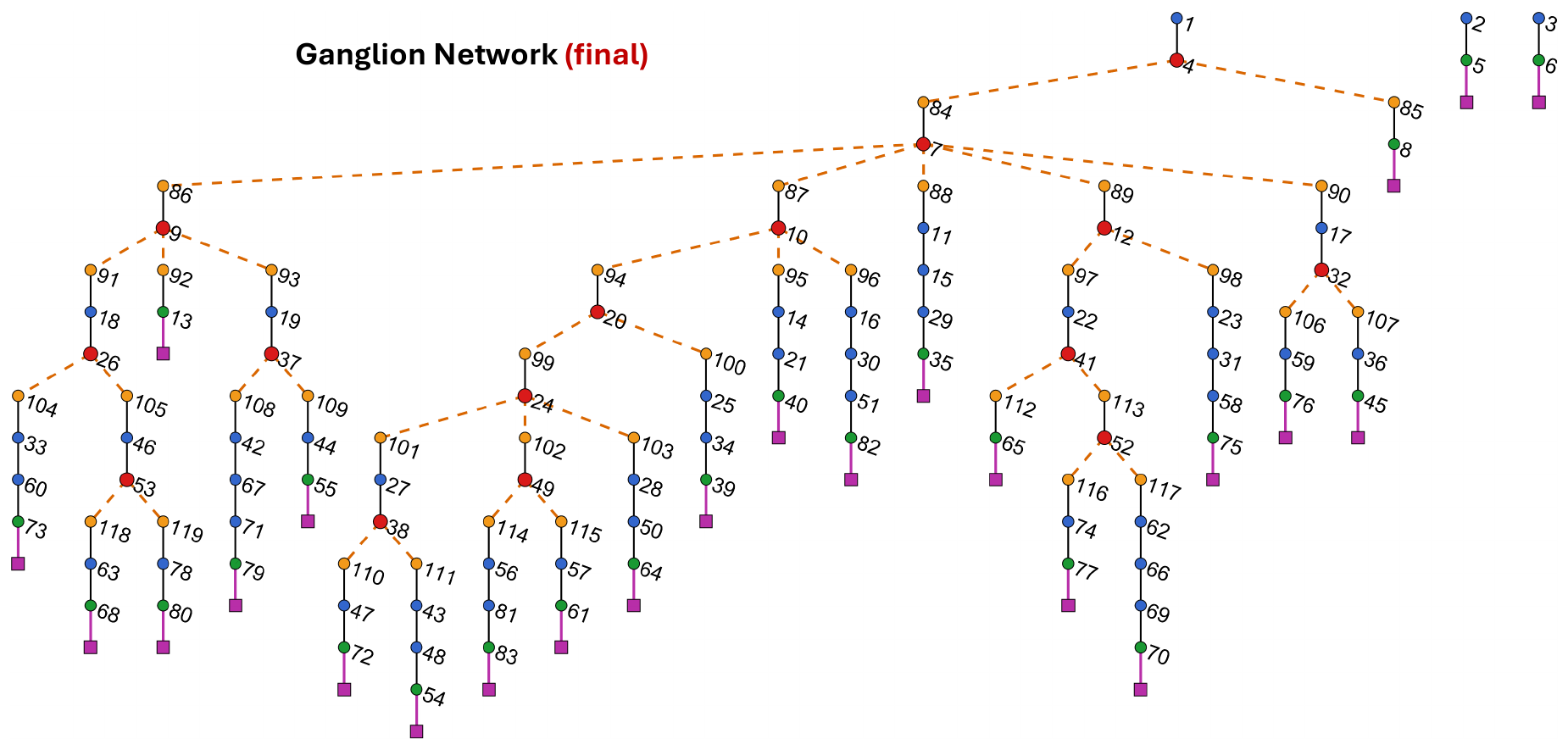}
\caption{Ganglion network of Fig.~\ref{fig:construct} after the adjustments of Section~\ref{sec:extract}. Leaf links (magenta) and terminal nodes (magenta squares) are appended below leaf nodes (green), and graph is coarsened with $f_\kappa\!=\!0.1$. Junction nodes are red, virtual nodes and virtual links orange, regular nodes blue.}
\label{fig:coarse}
\end{figure}

\noindent\textbf{Coarsening.}
Recall the initial graph is built by opening with every radius from $r\!=\!1$ to $r_{max}$, so a new node is created each time $r$ grows by one pixel. Therefore, every path down the tree from a root or virtual node to the next junction or leaf node passes through a sequence of regular nodes, which we call a \textit{chain}, whose consecutive nodes differ by one pixel in $r$. This spacing is set by the resolution of the image, not the geometry of the void space, and many of these nodes are redundant. The result is that the nonlinear curvature--volume relation $\kappa(V)$ along a chain is discretized too finely with piecewise linear interpolants given by Eq.~\ref{eq:intrpcrv}.
We thus proceed to coarsen the graph by removing a subset of the regular nodes using the line-simplification algorithm of Douglas and Peucker~\cite{douglas1973dp}, applied to $\kappa(V)$ along each chain. Doing so allows for taking larger time steps in GNM, as discussed in Section~\ref{sec:numerics}.
A regular node is removed only if the curvature from Eq.~\ref{eq:intrpcrv} across the longer link that replaces it, evaluated at the node's volume $V_i$, differs from the node's own $\kappa_i$ by less than a fraction $f_\kappa$ of the difference between the largest and smallest curvatures along the chain.
Junction, virtual, leaf, and terminal nodes are not removed, and $(\kappa_i, V_i, \mathbf{x}_i)$ at all retained nodes are unchanged.
 We use $f_\kappa\!=\!0.1$ here, which reduces nodes and links by $\sim$2 times with negligible impact on accuracy (Section~\ref{sec:disc}).
Fig.~\ref{fig:coarse} illustrates the initial graph of Fig.~\ref{fig:construct} after leaf links are appended (magenta) and the graph is coarsened.
\\

We conclude with a few remarks and definitions. Note that the entry curvature of a virtual node is always larger than the snap-off curvature of its parent junction node, capturing the well known hysteresis between invasion and snap off. The opening radius $r_j$ that created each junction node in the initial graph is preserved for later computations in Eq.~\ref{eq:AL}.
In GNM, each ganglion is represented by a point that lies somewhere on the extracted tree graph. A ganglion can reside on any link, except on virtual links (excluding their end nodes).
We define the \textit{state} of a ganglion by that ganglion's volume $V_b$ and the link it occupies. These two pieces of information also determine the ganglion's curvature $\kappa_b$ through Eqs.~\ref{eq:leafcrv} and \ref{eq:intrpcrv}, and the ganglion's spatial position through a nodal interpolation similar to Eq.~\ref{eq:intrpcrv}.
The extracted ganglion network is thus the appropriate phase space for multi-pore ganglia, as argued in Section~\ref{sec:intro}.

\subsection{Mean-field equations}
\label{sec:meanfield}

The GNM evolves a ganglion population through an effective-medium closure in the tradition of kinetic theories of Ostwald ripening~\cite{lifshitz1961orig, wagner1961orig, yu2023GRLkinetics, bueno2024theory, bueno2025theory}. Ganglia exchange mass with each other through a \textit{mean field} that represents the water and stores dissolved solute. The mean field is characterized by a single, spatially uniform solute mole fraction $X_m(t)$. Recall from Section~\ref{sec:extract} that the state of a ganglion $i$ consists of its volume $V_{b,i}$ and the link it occupies on the graph. Its interfacial curvature $\kappa_{b,i}$ follows from Eqs.~\ref{eq:leafcrv}--\ref{eq:intrpcrv}. Combining Eqs.~\ref{eq:henry} and \ref{eq:younglaplace}, the solute concentration adjacent to a ganglion is $X_{b,i}\!=\!(p_w - p_v + \sigma\kappa_{b,i})/H$.
Each ganglion exchanges mass with the mean field diffusively at the rate:
\begin{equation}
\rho_b \dd{V_{b,i}}{t} = \rho_w D_m \frac{\avg{A}}{\avg{L}}\left(X_m - X_{b,i}\right).
\label{eq:gangrate}
\end{equation}
Here, $\rho_b$ and $\rho_w$ are, respectively, the molar densities of the ganglion phase and water, $D_m$ is the diffusion coefficient of the solute, $\avg{A}$ is the mean cross-sectional area through which a ganglion exchanges solute mass with the surrounding water, and $\avg{L}$ is the mean spacing between the ganglia trapped in the porous medium.
Note that a ganglion whose concentration satisfies $X_{b,i}>X_m$ dissolves into the mean field, while a ganglion with $X_{b,i}<X_m$ grows.
Eq.~\ref{eq:gangrate} can also be written in terms of the mean-field curvature $\kappa_m\!=\!(H X_m - p_w + p_v)/\sigma$ to yield the form given by Eq.~\ref{eq:gnmclosure}.

To compute $\avg{A}$ and $\avg{L}$, we use:
\begin{equation}
\avg{A} = \frac{1}{n_j}\sum_{j=1}^{n_j} A_j,
\qquad
\avg{L} = \left(\frac{V_p}{n_G\,\phi\,g}\right)^{1/2}  \quad (2.5\mathrm{D}),
\qquad
\avg{L} = \left(\frac{V_p}{n_G\,\phi}\right)^{1/3}  \quad (3\mathrm{D}),
\label{eq:AL}
\end{equation}
where $n_j$ is the number of junctions in the graph, $V_p$ is the total pore volume of the porous medium, and $\phi$ is its porosity.
The expression for $\avg{A}$ is the arithmetic average of all throat cross-sectional areas $A_j$ over all junctions in the graph. The latter is set to $A_j\!=\!2 r_j g$ in 2.5D, and to $A_j\!=\!4r_j^2$ in 3D, where $r_j$ denotes the opening radius used to create junction node $j$ in the initial graph and $g$ is the out-of-plane gap thickness.
The expression for $\avg{L}$ assumes the $n_G$ ganglia are distributed uniformly across the domain. Notice Eq.~\ref{eq:gangrate} ignores the multiplicity of throats (coordination number) through which each ganglion exchanges mass with the mean field. This is deliberate given the extra, non-trivial image analysis it requires as well as other sources of uncertainty that get absorbed into $\avg{A}/\avg{L}$ in Eq.~\ref{eq:gangrate}. We thus aim to capture the order of magnitude of this multiplier, validated later.

To complete the formulation, we need an expression for $X_m$, which coupled with Eq.~\ref{eq:gangrate} describes how the ganglion population evolves. This extra expression is the balance of total moles $N_t$ for the ganglion species:
\begin{equation}
\dd{N_t}{t} = \dov{t}\left(\rho_w V_w X_m + \rho_b \sum_i V_{b,i}\right) = \rho_w D_m C_\partial \left(X_\partial - X_m\right),
\label{eq:meanfield}
\end{equation}
where $V_w\!=\!V_p - \sum_i V_{b,i}$ is the volume of water. The first term inside the parentheses corresponds to the dissolved solute and the second to the ganglion phase. The right side describes the diffusive exchange with an external reservoir of concentration $X_\partial$ through a boundary conductance $C_\partial$. We compute $C_\partial$ as the sum of all area-to-length ratios of throats intersecting the domain boundary.
To solve Eqs.~\ref{eq:gangrate}--\ref{eq:meanfield}, we need the initial ganglion states (thus $V_{b,i}$), the initial $X_m$, and reservoir concentration $X_\partial$. Note that water in contact with a flat interface contains dissolved concentration $X_{mo}\!=\!(p_w - p_v)/H$. Here, we consider three evolution scenarios of a ganglion population: (1) \textit{ripening}, set by $X_m(t\!=\!0)\!=\!X_\partial\!=\!X_{mo}$; (2) \textit{dissolution}, set by $X_m(t\!=\!0)\!=\!X_\partial\!<\!X_{mo}$; and (3) \textit{growth}, set by $X_m(t\!=\!0)\!=\!X_\partial\!>\!X_{mo}$.
These scenarios represent a population suddenly placed in contact with a large body of water of concentration $X_\partial$.
\\

\noindent\textbf{Remark 1.} In existing theories of Ostwald ripening, the water carries no storage capacity for dissolved solute~\cite{lifshitz1961orig, wagner1961orig, yu2023GRLkinetics, bueno2024theory, bueno2023PNM}.
Formally, the storage term $\rho_w V_w X_m$ and the boundary-exchange term in Eq.~\ref{eq:meanfield} are neglected, leading to:
\begin{equation}
\sum_i \dd{V_{b,i}}{t} = 0.
\label{eq:nostor1}
\end{equation}
Substituting Eq.~\ref{eq:gangrate} into Eq.~\ref{eq:nostor1} and simplifying yields:
\begin{equation}
X_m = \frac{1}{n_G}\sum_i X_{b,i} = \frac{p_w - p_v + \sigma\bar{\kappa}}{H},
\qquad \bar{\kappa} = \frac{1}{n_G}\sum_i \kappa_{b,i},
\label{eq:nostor2}
\end{equation}
where the second equality follows from $X_{b,i}\!=\!(p_w - p_v + \sigma\kappa_{b,i})/H$ and $\bar{\kappa}$ is the population-mean curvature. In words, the mean field equilibrates instantaneously to the average curvature of ganglia. By contrast, GNM captures the solute stored within the water through Eq.~\ref{eq:meanfield} allowing for modeling ganglion evolutions beyond Ostwald ripening.

\subsection{Rules of ganglion motion}
\label{sec:rules}

\begin{figure}[b!]
\centering
\includegraphics[width=\linewidth]{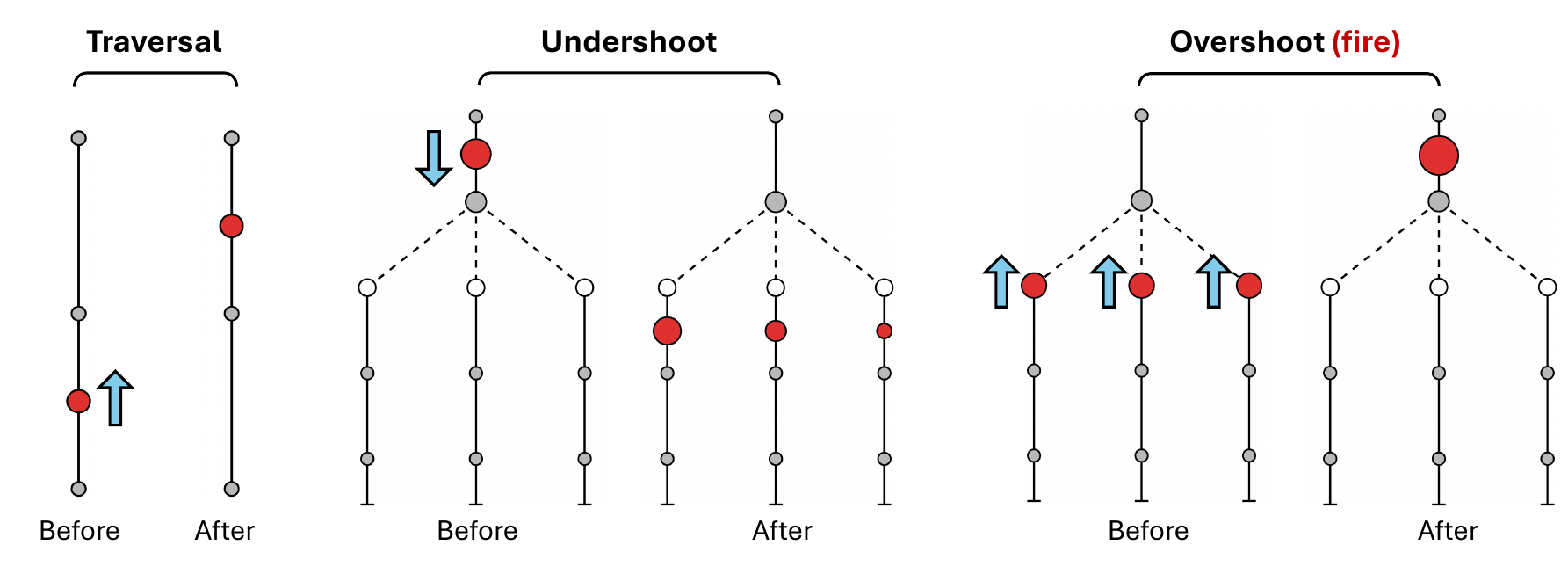}
\caption{Three rules of ganglion motion, illustrated through a three-way junction with before and after pictures. Ganglia are shown as red circles whose sizes encode volume, virtual nodes as open circles, virtual links as dashed lines, and all other regular nodes as gray circles. Blue arrows indicate the direction a ganglion is moving. In a \textit{traversal} (left), a ganglion reaching a regular node simply moves onto the only other regular link attached to the node. In an \textit{undershoot} (middle), a ganglion that shrinks past a junction node breaks up into one fragment per branch, with volumes proportional to the virtual-node volumes. In an \textit{overshoot} (right), if a ganglion sits at every virtual node of the junction, any further growth causes a merger into one larger ganglion sitting on the parent link of the junction node. We call this event a \textit{fire}, which is a special kind of overshoot.}
\label{fig:rules}
\end{figure}

\begin{figure}[t!]
\centering
\includegraphics[width=\linewidth]{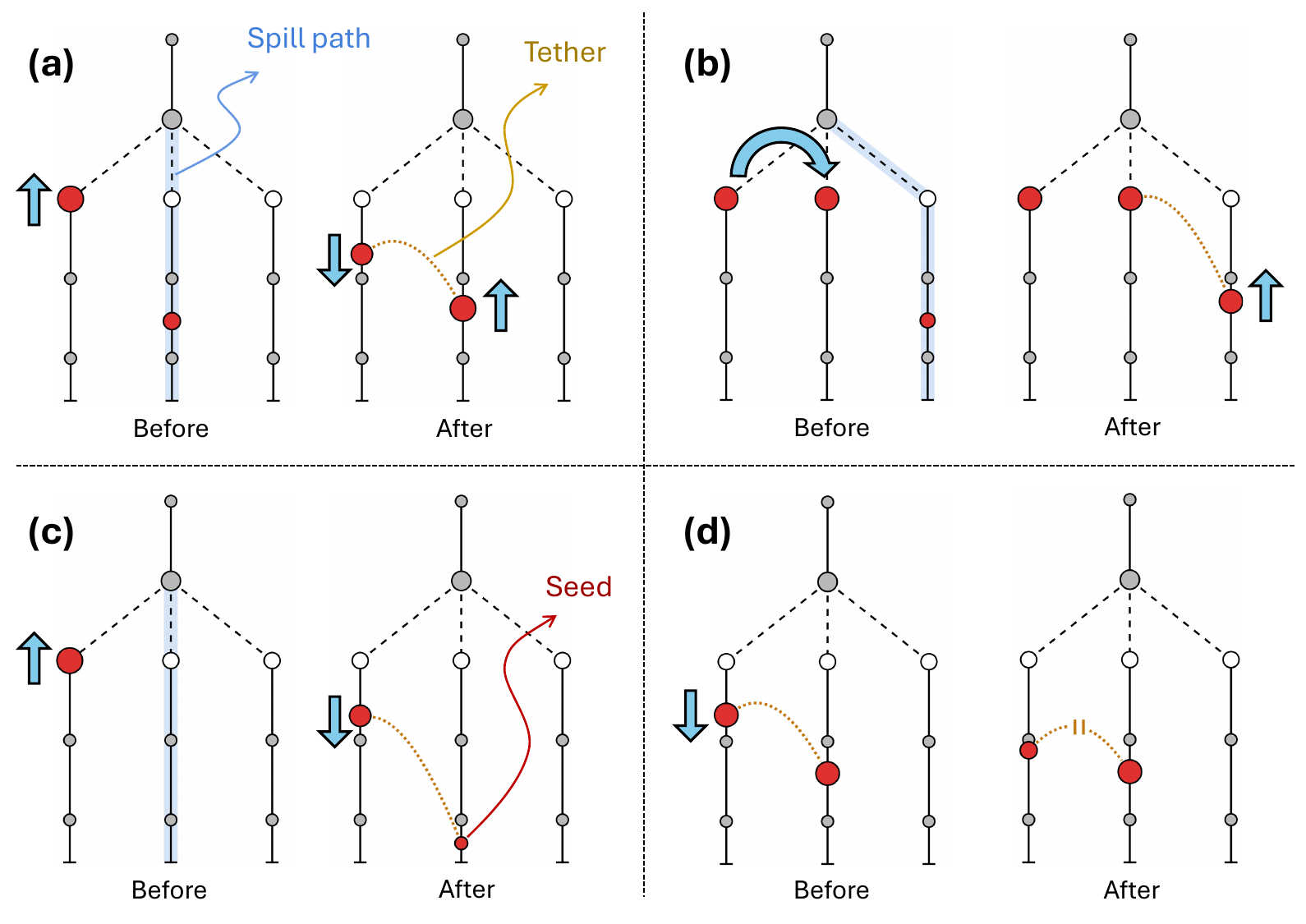}
\caption{Overshoot events when not all virtual nodes of a junction are occupied by a ganglion, illustrated through a three-way junction with before and after pictures. Plotting conventions follow those of Fig.~\ref{fig:rules}. The shaded branch marks the \textit{spill path} and the orange dotted line is a \textit{tether}. (a) The growing ganglion passes its excess volume into the spill path and onto the ganglion sitting on it. The receiving ganglion grows as a result, the spilling ganglion slides down its link, and the two become tethered. (b) If the receiving ganglion overshoots at the same junction as the spiller, the receiver becomes a spiller itself. (c) If the spill path is vacant, the spilled volume seeds a new ganglion on the leaf link, tethered to the spiller. (d) A tether is severed when the curvature of either ganglion in the pair (here left) drops below the snap-off curvature of the anchoring junction node.}
\label{fig:spill}
\end{figure}

Consider now a population of ganglia occupying the porous medium.
Each ganglion's state, or position on the tree graph, is determined by its volume $V_{b,i}$ and the regular link $\ell$ it occupies, which in turn fixes its curvature $\kappa_{b,i}$ through Eqs.~\ref{eq:leafcrv}--\ref{eq:intrpcrv} as already discussed.
Moreover, each ganglion's state satisfies the following self-evident property:
\\

\noindent\textbf{Antichain property.} If ganglion $\mathrm{A}$ occupies link $\ell$, no other ganglion $\mathrm{B}$ is allowed to reside anywhere within the subtree that lies underneath $\ell$. Violation would imply $\mathrm{A}$ and $\mathrm{B}$ overlap spatially, because $\mathrm{B}$ is a fragment of $\mathrm{A}$.
\\

To evolve the population in time on the graph, Eqs.~\ref{eq:gangrate} and \ref{eq:meanfield} are discretized and advanced in time in staggered fashion. Each time step, $V_{b,i}$ is incremented for all $i$ by a forward Euler step of Eq.~\ref{eq:gangrate} at the current mean-field concentration $X_m$. Then, $X_m$ is updated implicitly through Eq.~\ref{eq:meanfield}. As each $V_{b,i}$ evolves, the corresponding ganglion $i$ moves along its link $\ell$, with growth carrying it upward toward the parent node and dissolution carrying it downward toward the child node. Since time steps are finite, a ganglion may reach one of the two nodes with volume change to spare. At that instant, one of three rules is applied: \textit{traversal}, \textit{undershoot}, or \textit{overshoot}. Below, we describe each rule in turn, with Figs.~\ref{fig:rules}--\ref{fig:spill} illustrating them on simple graphs. Ganglia are drawn as red circles whose sizes encode volume. The rules assume $V_{b,i}$ is already incremented via Eq.~\ref{eq:gangrate}, and dictate how the link index $\ell$ and $\kappa_{b,i}$ must be updated.
The rules are applied iteratively and recursively until every ganglion resides in the interior of some link.
\\

\noindent\textbf{Traversal.} If the node reached is a regular node, the ganglion moves onto the only other link $\ell^{\ast}$ connected to that node (Fig.~\ref{fig:rules}). The move leaves $V_{b,i}$ intact while causing $\kappa_{b,i}$ to change continuously across the node.
The ganglion is then re-examined at $\ell^{\ast}$, to see if $V_{b,i}$ fits between the new link's nodal volumes. If not, the appropriate rule is applied again.
Note that repeated traversals are how ganglia move along the non-branching chains of the tree graph.
\\

\noindent\textbf{Undershoot.} If the ganglion is shrinking and the (child) node of $\ell$ reached is a junction $j$ with $n$ virtual child links, the ganglion breaks up into $n$ fragments (Fig.~\ref{fig:rules}). Each fragment $k\!\in\!\{1,\cdots,n\}$ is assigned a fraction of the ganglion's volume $V_{b,i}$ in proportion to the virtual-node volumes $V_{v,k}$ under the junction, i.e., $V_{b,k}\!=\!\left(V_{v,k}/\sum_m V_{v,m}\right)V_{b,i}$.
The fragment is placed on the regular link $\ell_k$ under the virtual node $v_k$, then re-examined to see if $V_{b,k}$ fits on $\ell_k$. If not, the appropriate rule is invoked again. Hence, fragment $k$ resides somewhere on the subtree rooted at the virtual node $v_k$, henceforth referred to as \textit{branch} $k$ of junction $j$.
Notice fragments skip virtual links, as no ganglion can occupy them.
A special case arises if $\ell$ is a leaf link: the ganglion reaching the child node has zero volume and is removed from the graph (Remark 3). Physically, undershoot represents a ganglion's fragmentation by snap off at its interior throats.
\\

\noindent\textbf{Overshoot.} If the ganglion is growing and the (parent) node of $\ell$ reached is a virtual node $v_k$ of some junction $j$ with $n$ branches, the ganglion is said to have \textit{filled} branch $k$ of that junction.
The \textit{excess} volume $V_{e,i}\!=\!V_{b,i}\!-\!V_{v,k}$ must be passed into one of the non-filled branches $m\!\neq\!k$ of $j$, physically corresponding to the capillary invasion of the junction's throat(s).
But if all other branches of $j$ are filled too, meaning a ganglion resides at their virtual nodes, then all ganglia under $j$ are removed and replaced by one ganglion on the parent link $\ell_p$ of the junction node. The volume of the merged ganglion equals the sum of the ganglion volumes removed (including the excess). We refer to this event as a \textit{fire} (Fig.~\ref{fig:rules}), and depending on whether the merged ganglion fits on $\ell_p$, the appropriate rule is invoked again.

The more common scenario is when the junction has at least one non-filled branch $m\!\neq\!k$, in which case the ganglion at branch $k$ is said to \textit{spill} over into branch $m$. Physically, the spill represents a capillary invasion of the ganglion at branch $k$ into pores represented by the virtual node $v_m$. During a spill, a portion of the ganglion's volume is passed into branch $m$, which includes the excess $V_{e,i}$ plus an extra portion $\delta V_{b,i}$ that causes the spilling ganglion to \textit{slide down} its link $\ell$.
The $\delta V_{b,i}$ aims to capture the experimentally observed fact that an invading ganglion could empty much of its volume into a neighboring pore, provided it has enough room~\cite{mehmani2022AWR}.
To compute it, we use:
\begin{equation}
\delta V_{b,i} = \min\left( V_{v,k} - V_{s,k},\ \max\left( V_{v,m} - V_{e,i} - \sum_{q \,\in\, B_m} V_{b,q},\ 0 \right) \right),
\label{eq:slide}
\end{equation}
where $V_{s,k}$ is called the \textit{snap-off volume} of branch $k$, and the sum is over all ganglia $q$ that lie on branch $m$---denoted by $B_m$.
To compute $V_{s,k}$, we move down the chain of links that sits underneath the virtual node $v_k$ until the interpolated curvature via Eq.~\ref{eq:intrpcrv} equals the junction's snap-off curvature $\kappa_s$. The corresponding volume is then $V_{s,k}$.
But if equality is not achieved before a lower junction node or a leaf node $i^\ast$ is reached, then $V_{s,k}\!=\!V_{i^\ast}$.
Eq.~\ref{eq:slide} takes the minimum of two values: (1) slid-down volume of the spilling ganglion, and (2) room available in branch $m$ to receive the spill.

With the spilled volume $V_{e,i}+\delta V_{b,i}$ computed, we must now select a unique path---a sequence of serially connected links---within branch $m$ to dump this volume into.
We call this path the \textit{spill path}, and its selection corresponds physically to deciding which of the neighboring pore(s) to invade. From Section~\ref{sec:extract}, recall the throat(s) at junction $j$ have identical entry curvature. Our policy is to pick the closest pore by Euclidean distance to one of the pores occupied by the spilling ganglion. Since leaf nodes represent pores, we pick the leaf node $f_m$ of branch $m$ that is closest to the set of all leaf nodes of branch $k$.
Recall each node $i$ in the graph has a stored position $\mathbf{x}_i$.
Once selected, we walk up the tree from $f_m$ to the virtual node $v_m$, with the route traveled defining the spill path.
We next dump $V_{e,i}+\delta V_{b,i}$ into this spill path, which gives rise to two cases: (1) If another ganglion already lies somewhere on the spill path, the dumped volume is absorbed by it, causing it to move up the tree (Fig.~\ref{fig:spill}a).
If the absorbing ganglion no longer fits on its current link, the appropriate rule is invoked again, which may result in multiple cascading spill events (Fig.~\ref{fig:spill}b); (2) If the spill path is empty, with no ganglion occupying it, the spilled volume is used to spawn a new ganglion, called a \textit{seed}, on the leaf link (Fig.~\ref{fig:spill}c).
Spawning requires a minimum amount of spilled volume, elaborated in Remark 3.

Since a spill event represents the invasion of the growing ganglion into a neighboring pore, the spilling ganglion and the spawned or absorbing ganglion become connected, as two parts of a single non-wetting phase body.
In GNM, this connection is recorded by declaring that the pair is \textit{tethered} around junction $j$ (Fig.~\ref{fig:spill}a--c).
A tether is severed when the curvature of either part drops below the anchoring junction's snap-off curvature $\kappa_s$ (Section~\ref{sec:extract}), at which point the body snaps off at the connecting throat(s) and the two parts disconnect into independent ganglia (Fig.~\ref{fig:spill}d).
\\

\noindent\textbf{Remark 2.} Consider an overshooting ganglion on branch $k$ of junction $j$, spilling into branch $m$. If the ganglion slides down (Fig.~\ref{fig:spill}a) \textit{and} the receiving branch overshoots at $j$, then the above approach for selecting the spill path based on the proximity of leaf nodes will cause a second spill from branch $m$ back into branch $k$ (which now has room)---triggering an infinite loop.
To break the loop, we exclude branch $k$ from the second spill-path selection, ensuring a cascade similar to Fig.~\ref{fig:spill}b.
If further overshoots occur in the same cascade, we exclude all previously overshot branches from the next spill-path selection. If all branches get excluded with one branch still overshooting, we purge the exclusion list and start afresh. This resolves the infinite loop and causes an eventual \textit{fire} if all branches are filled.
\\

\noindent\textbf{Remark 3.} Leaf links are where shrinking ganglia vanish and new ones (seeds) are spawned.
However, tiny spherical (3D) or disc-shaped (2.5D) bubbles on leaf links have very high curvature $\kappa_b$ (Eq.~\ref{eq:leafcrv}), which affect the numerical accuracy of the mean-field concentration $X_m$ through Eqs.~\ref{eq:gangrate} and~\ref{eq:meanfield}.
Moreover, ganglia born with $\kappa_b$ larger than the mean-field curvature $\kappa_m\!=\!(H X_m - p_w + p_v)/\sigma$ immediately dissolve back into water, artificially restricting the time-step size.
To address both, we: (1) remove a shrinking ganglion on a leaf link if its volume drops below a small threshold $V_{min}$ (value given in Section~\ref{sec:numerics});
and (2) spawn a ganglion, due to a spill, only if its volume exceeds $V_{min}$ and its would-be curvature on the leaf link is below $\kappa_m$.
To prevent mass gain/loss, we store the post-vanish residual volume of the ganglion and the pre-spawn volume of the seed inside a container $V_{a,f}$ at each leaf link $\ell_f$.
The accumulated volume in $V_{a,f}$ is counted towards the total ganglion-phase volume but is excluded from mean-field calculations.
An exception is made if $V_{a,f}$ reaches the leaf-node volume, in which case a seed is spawned regardless of its curvature.

\subsection{Numerical solution}
\label{sec:numerics}

Algorithm~\ref{alg:gnm} summarizes the pseudocode of GNM for evolving a partially miscible ganglion population inside a porous medium. The algorithm consists of a time loop, within which the following steps are executed: (1) Given each ganglion's state $(V_{b,i},\ell_i)$, its curvature $\kappa_{b,i}$ is computed via Eqs.~\ref{eq:leafcrv}--\ref{eq:intrpcrv}, followed by its interface-adjacent concentration $X_{b,i}$ via Henry's law; (2) substituting $X_{b,i}$ into Eq.~\ref{eq:gangrate} yields the rate of volume change $\dot{V}_{b,i}$ for each ganglion; (3) the rates $\dot{V}_{b,i}$ are then used to select an adaptive time step $\delta t$, as described below; (4) ganglion volumes are incremented explicitly by $\delta t\,\dot{V}_{b,i}$; (5) next, ganglion links, $\ell_i$, are updated by moving the ganglia on the extracted tree graph, and applying the appropriate rules described in Section~\ref{sec:rules} and summarized in Algorithm~\ref{alg:rules}.
The rules may cause existing ganglia to fragment, merge, vanish, or change size, and new ones to spawn; finally, (6) the mean-field concentration $X_m$ is updated implicitly via Eq.~\ref{eq:meanfield} using the new ganglion states.
The cost of GNM is controlled by Step 5, tracking ganglia on the graph, which scales with population size but is independent of domain size (unlike PNM).\looseness=-1

Algorithm~\ref{alg:rules} updates the ganglion states $(V_{b,i},\ell_i)$, including any tethers between them, given an input state where the volumes $V_{b,i}$ alone are perturbed.
The procedure is iterative and stops when every ganglion resides in the interior of some link.
Let $V^{c}_{\ell}$ and $V^{p}_{\ell}$ denote the volumes stored at the child and parent nodes of link $\ell$, respectively.
For a ganglion satisfying $V_{b,i}\!<\!V^{c}_{\ell}$, we check if the traversal or undershoot rules apply. If $\ell$ corresponds to a leaf link, the ganglion vanishes.
Similarly, for a ganglion satisfying $V_{b,i}\!>\!V^{p}_{\ell}$, we check if the traversal or overshoot rules apply, the latter potentially leading to a cascading spill.
Tethers are checked for snap off after the population settles.

\floatstyle{plain}
\restylefloat{algorithm}
\providecommand{\algwidth}{0.80\textwidth}
\providecommand{\algindent}{0.07\textwidth}
\providecommand{\algcom}[1]{\hfill\makebox[0.26\linewidth][l]{$\triangleright$~#1}}
\providecommand{\alghdind}{0.03\textwidth}
\providecommand{\alghead}[1]{%
  \hrule height 0.9pt\vspace{3pt}%
  \noindent\hspace*{\alghdind}\textbf{Algorithm \thealgorithm.} #1\vspace{3pt}%
  \hrule\vspace{0.9em}}

\begin{algorithm}[t!]
\centering
\begin{minipage}{\algwidth}
\small
\refstepcounter{algorithm}\label{alg:gnm}
\alghead{Ganglion network model}
\hspace*{\algindent}%
\begin{minipage}{\dimexpr\linewidth-\algindent\relax}
\begin{algorithmic}
\State \textbf{Input:} $\{(V_{b,i},\ell_i)\}$, $X_m$, tethers, $T$
\State \textbf{Output:} $\{(V_{b,i},\ell_i)\}$, $X_m$, tethers
\State $t=0$; initialize ganglion states
\State \textbf{Do} while $t<T$
\State \quad Compute $\kappa_{b,i}$ from Eqs.~\ref{eq:leafcrv}--\ref{eq:intrpcrv}
\State \quad Compute $\dot{V}_{b,i}$ from Eq.~\ref{eq:gangrate}
\State \quad Compute $\delta t$ from Eq.~\ref{eq:dt} \algcom{adaptive}
\State \quad Update $V_{b,i} \leftarrow V_{b,i} + \delta t\,\dot{V}_{b,i}$ \algcom{explicit}
\State \quad Apply the rules of motion \algcom{Algorithm~\ref{alg:rules}}
\State \quad Solve Eq.~\ref{eq:meanfield} for $X_m$ \algcom{implicit}
\State \quad $t \leftarrow t + \delta t$
\State \textbf{End do}
\end{algorithmic}
\end{minipage}
\vspace{0.6em}
\hrule height 0.9pt
\end{minipage}
\end{algorithm}

\begin{algorithm}[t!]
\centering
\begin{minipage}{\algwidth}
\small
\refstepcounter{algorithm}\label{alg:rules}
\alghead{Rules of motion}
\hspace*{\algindent}%
\begin{minipage}{\dimexpr\linewidth-\algindent\relax}
\begin{algorithmic}
\State \textbf{Input:} $\{(V_{b,i},\ell_i)\}$, tethers
\State \textbf{Output:} $\{(V_{b,i},\ell_i)\}$, tethers
\State \textbf{Repeat}
\State \quad \textbf{Do} for every ganglion with $V_{b,i}\!<\!V^{c}_{\ell_i}$
\State \quad \quad Apply traversal if the child node is regular
\State \quad \quad Apply undershoot if the child node is a junction
\State \quad \quad Remove the ganglion if $\ell_i$ is a leaf link \algcom{Remark 3}
\State \quad \textbf{End do}
\State \quad \textbf{Do} for every ganglion with $V_{b,i}\!>\!V^{p}_{\ell_i}$
\State \quad \quad Apply traversal if the parent node is regular
\State \quad \quad Apply overshoot if the parent node is virtual \algcom{fire or spill}
\State \quad \textbf{End do}
\State \textbf{Until} every ganglion resides in the interior of its link
\State Sever every tether whose parts fall below $\kappa_s$ \algcom{snap off}
\end{algorithmic}
\end{minipage}
\vspace{0.6em}
\hrule height 0.9pt
\end{minipage}
\end{algorithm}

The time step is chosen adaptively, controlled by the first ganglion to reach a node.
Defining $\Delta V_{\ell}\!=\!V^{p}_{\ell}-V^{c}_{\ell}$ and $\Delta\kappa_{\ell}\!=\!|\kappa^{p}_{\ell}-\kappa^{c}_{\ell}|$ as the volume and curvature spans of link $\ell$, respectively, the time step $\delta t$ is set by:
\begin{equation}
\delta t = \min_i \delta t_i,
\qquad
\delta t_i =
\begin{cases}
\max\left(\tau_i,\ \delta t^{flr}_i\right), & \tau_i = \delta t^{nod}_i\\[2pt]
\tau_i, & \text{otherwise}
\end{cases},
\qquad
\tau_i = \min\left(\delta t^{nod}_i,\ \delta t^{cap}_i,\ \delta t^{equ}_i\right),
\label{eq:dt}
\end{equation}
where
\begin{equation}
\delta t^{nod}_i = \begin{cases}
\left(V^{p}_{\ell}-V_{b,i}\right)\,/\,|\dot{V}_{b,i}|, & \dot{V}_{b,i}>0\\[2pt]
\left(V_{b,i}-V^{c}_{\ell}\right)\,/\,|\dot{V}_{b,i}|, & \dot{V}_{b,i}<0
\end{cases},
\quad
\delta t^{flr}_i = \frac{f_{min}\Delta V_{\ell}}{|\dot{V}_{b,i}|},
\quad
\delta t^{cap}_i = \frac{f_{max}\Delta V_{\ell}}{|\dot{V}_{b,i}|},
\quad
\delta t^{equ}_i = f_{eq}\frac{|\kappa_m-\kappa_{b,i}|\,\Delta V_{\ell}}{\Delta\kappa_{\ell}\,|\dot{V}_{b,i}|},
\label{eq:dtcand}
\end{equation}
and the index $i$ runs over all ganglia.
In Eq.~\ref{eq:dtcand}, the first expression, $\delta t^{nod}_i$, is the time for ganglion $i$ to reach the node it is approaching.
The next two bound the volume change a ganglion may undergo in one step to between $f_{min}\Delta V_{\ell}$ and $f_{max}\Delta V_{\ell}$.
The floor $\delta t^{flr}_i$ ensures that a ganglion very close to its approaching node is carried slightly past that node, so one of the rules of Section~\ref{sec:rules} applies.
If absent, the ganglion could land exactly on the node resulting in the next time step to become zero.
The last expression, $\delta t^{equ}_i$, prevents a step from carrying $\kappa_{b,i}$ across the mean-field curvature $\kappa_m$, which would flip the ganglion between growth and dissolution and produce spurious oscillations about equilibrium.
As the population approaches equilibrium, the driving force $\kappa_m-\kappa_{b,i}$ (thus $\dot{V}_{b,i}$) drops, and $\delta t$ grows adaptively.

A leaf link is a special case, as its child node stores $V^{c}_{\ell}\!=\!0$ and $\kappa^{c}_{\ell}\!=\!\infty$, forcing $\delta t^{equ}_{i}\!=\!0$. We override this by setting $\delta t^{equ}_{i}\!=\!\infty$ on leaf links to avoid the time step $\delta t$ from becoming zero.
Conversely, we set $\smash{\delta t^{flr}_{i}\!=\!0}$ for shrinking ganglia on leaf links to ensure their approach towards the vanishing threshold $V_{min}$ is gradual and controlled (Remark 3).

We set $f_{min}\!=\!5\!\times\!10^{-4}$, $f_{max}\!=\!2\!\times\!10^{-3}$, and $f_{eq}\!=\!1/2$ herein. The vanishing threshold in Remark 3 is also set to:
\begin{equation}
V_{min} = \left(\frac{\Delta x}{2}\right)^{D} g \quad (2.5\mathrm{D}),
\qquad
V_{min} = \left(\frac{\Delta x}{2}\right)^{D} \quad (3\mathrm{D}),
\label{eq:vmin}
\end{equation}
where $\Delta x$ is the side length of one image voxel and $g$ is the out-of-plane gap thickness in 2.5D.
Notice this is $2^{-D}$ times the volume of a single voxel, so a ganglion is removed once it can no longer be resolved on the image.

\subsection{Visualization}
\label{sec:visualize}

\begin{figure}[t!]
\centering
\includegraphics[width=0.8\linewidth]{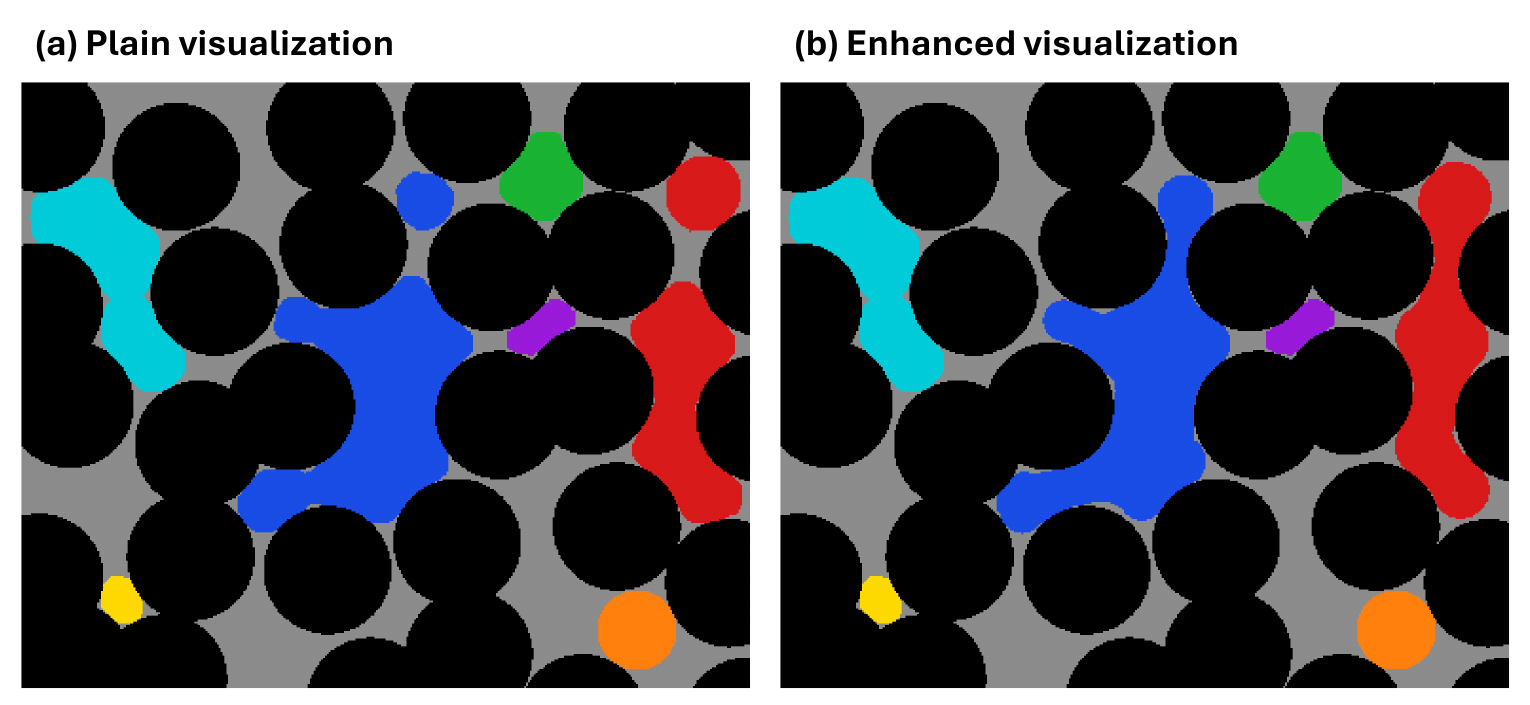}
\caption{Visualization of a ganglion population on the porous medium of Fig.~\ref{fig:construct}, with each body shown in a different color. (a) Plain rendering, in which each ganglion is drawn using the image voxels stored at the tree nodes straddling the occupied link. A tethered body spans more than one branch of a junction and therefore appears as disconnected patches sharing a color. (b) Skeleton-enhanced rendering, in which the patches of a tethered body are bridged along the medial axis of the void space, restoring the body's physical connectivity while preserving its total volume.}
\label{fig:viz}
\end{figure}

A useful feature of GNM is that the state of a ganglion on the tree graph maps directly back onto the pore-scale image. Recall from Section~\ref{sec:extract} that each node of the graph is a connected component of the opened image, with the set of image voxels known. A ganglion residing on a link can thus be rendered by interpolating between the voxel sets of the two nodes that straddle it, in proportion to its volume, and a population is rendered by repeating this for every ganglion (Fig.~\ref{fig:viz}a). The mapping is faithful to the underlying microstructure because the voxel sets come from the image itself rather than any geometric idealization.
This permits direct visual comparison against experiments.

The above plain visualization, however, comes with an artifact.
A tethered body on the graph corresponds physically to a connected region of the non-wetting phase, but consists of two ``ganglia'' residing on different branches of a junction. When rendered, these ganglia appear as disconnected patches as shown in Fig.~\ref{fig:viz}a.
To restore physical connectivity of tethered bodies, we introduce an enhanced visualization (Fig.~\ref{fig:viz}b) that bridges the patches along the medial axis of the void space.
For each tether, the shortest skeleton path joining the two patches is found, then the void is filled along it by drawing the maximum inscribed disc or sphere at each point of the path.
The resulting voxel set is then trimmed or extended at free ends of the skeleton until the rendered volume matches the body's volume.
We use this enhanced rendering throughout Section~\ref{sec:results}, while noting that it is purely cosmetic and has no impact on GNM's computations.
A similar enhanced rendering was proposed for the image-based pore-network model (iPNM)~\cite{laku2026ipnm}, which is also parameterized by the pore morphology method. In Section~\ref{sec:results}, we use iPNM to validate GNM.

\section{Validation set}
\label{sec:valid}

The proposed GNM rests on two approximations, which we test in Section~\ref{sec:results}. First, the mean field assumes that the dissolved concentration $X_m$ is spatially uniform in the wetting phase. Second, the rules of ganglion motion on the tree graph only approximate how ganglia actually evolve inside a porous medium. Otherwise, GNM simplifies neither the geometry of the microstructure nor that of the ganglia occupying it. To validate GNM, we compare its predictions to iPNM~\cite{laku2026ipnm}, which explicitly resolves the concentration field, the mass transfer between individual ganglia, and their spatial, geometric, and topological evolution. In~\cite{laku2026ipnm}, iPNM was validated against microfluidic experiments of Ostwald ripening of hydrogen ganglia. The version used here further conserves mass to machine precision (\ref{app:ipnm}).\looseness=-1

\begin{figure}[b!]
\centering
\includegraphics[width=0.8\linewidth]{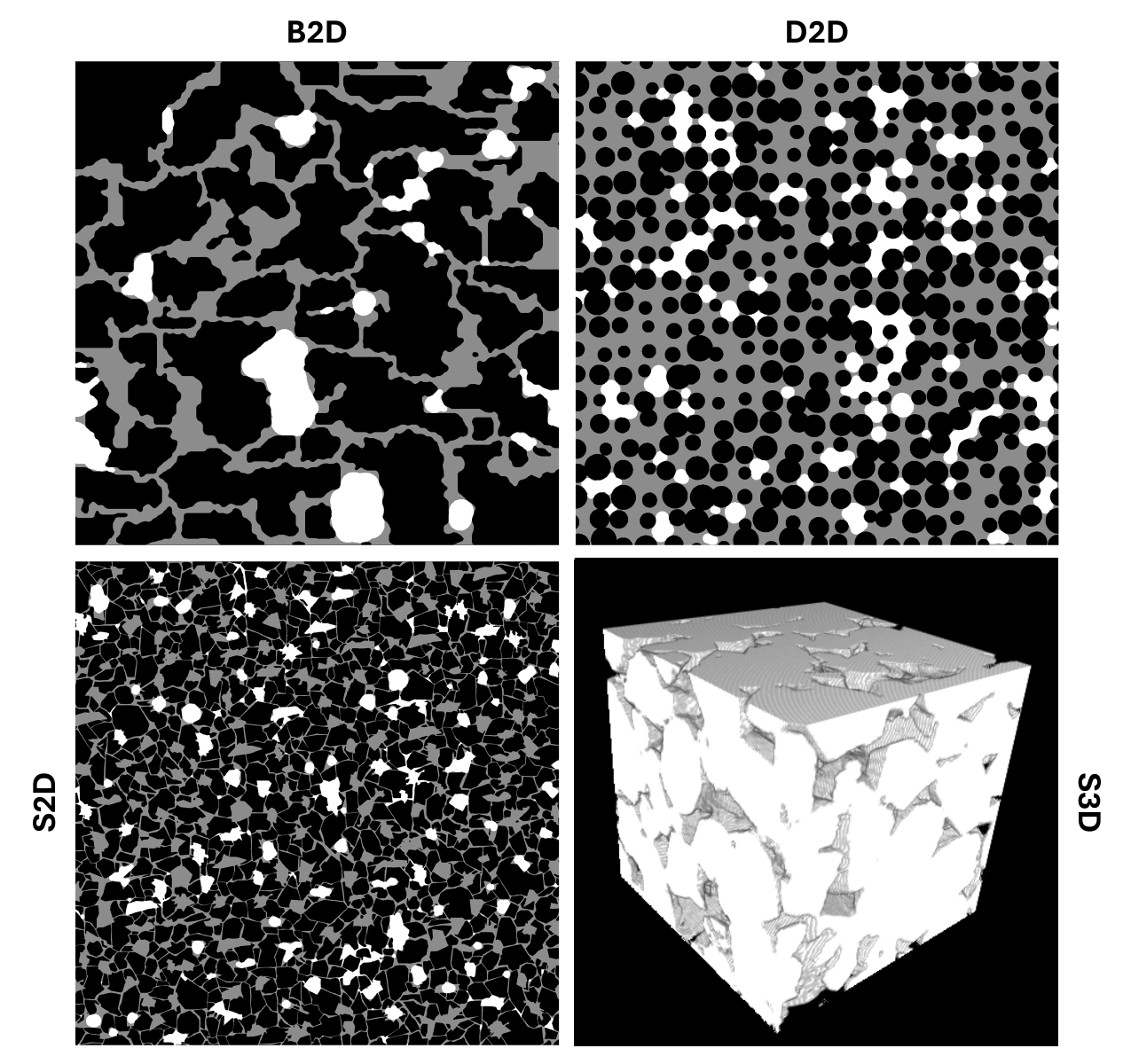}
\caption{The porous media used to validate GNM against iPNM. The domains consist of: (1) B2D, modified after a 2D cross section of Berea sandstone; (2) D2D, a synthetically generated polydisperse disc pack; (3) S2D, modified after a Canadian sandstone; and (4) S3D, modified after a 3D X-ray $\mu$CT image of a sandstone. In the first three, the configuration of ganglia at $t\!=\!0$ is also shown in white, with solid/void in black/gray.}
\label{fig:validate}
\end{figure}

We consider three 2.5D (planar with out-of-plane thickness) porous domains and one 3D domain, shown in Fig.~\ref{fig:validate}, with different microstructures: (1) \textit{B2D}, modified after a 2D cross section of Berea sandstone~\cite{boek2010lberea}; (2) \textit{D2D}, a synthetically generated polydisperse disc pack; (3) \textit{S2D}, modified after the micromodel of Salehpour et al.~\cite{salehpour2025micro}, itself patterned from a 2D X-ray CT image of a Canadian sandstone; and (4) \textit{S3D}, modified after a 3D X-ray $\mu$CT image of a sandstone~\cite{dong2009Sandstone} (``S5'' sample therein).
Table~\ref{tab:setup} lists each domain's dimensions, image size, voxel size ($\Delta x$), gap thickness ($g$), porosity ($\phi$), node and link counts in GNM, and pore and throat counts in iPNM.
The ganglion network of each domain is coarsened with $f_\kappa\!=\!0.1$ following Section~\ref{sec:extract}, and the node and link counts listed are those of the coarsened graph.
Since GNM and iPNM are both parameterized by the pore-morphology method, ganglia can be mapped onto the original pore-scale image without geometric simplification. This, in turn, allows for one-to-one visual comparison of the spatial configuration of ganglia through time.
The initial configuration at $t\!=\!0$ in each domain is shown in Fig.~\ref{fig:validate}, obtained by randomly placing ganglia on the GNM graph then mapping it to the image. The image is then used to map ganglia onto pores in iPNM, obtained from watershed segmentation of the microstructure. Thus, GNM and iPNM begin from identical initial conditions.
The ganglion count ($n_G$) at $t\!=\!0$ is also listed in Table~\ref{tab:setup} for each domain.\looseness=-1

We consider the three evolution scenarios defined in Section~\ref{sec:meanfield}, set through the ratio of the initial, spatially uniform water concentration $X_m^0$ to the flat-interface value $X_{mo}$:  (1) \textit{Ostwald ripening} ($X_m^0/X_{mo}\!=\!1$), (2) \textit{dissolution} ($X_m^0/X_{mo}\!=\!0.1$), and (3) \textit{growth} ($X_m^0/X_{mo}\!=\!10$). In GNM, the reservoir (or boundary) concentration $X_\partial$ is set equal to the initial value $X_m^0$ in all cases. In iPNM, this is equivalent to imposing $X_\partial\!=\!X_m^0$ as a boundary condition.
Ripening is the exception, where no-flux conditions are imposed at domain boundaries in iPNM, and equivalently $C_\partial\!=\!0$ (Eq.~\ref{eq:meanfield}) is set in GNM.
Ripening simulations, therefore, correspond to closed systems where the only mass transfer occurs between ganglia.
Table~\ref{tab:setup} further lists the fluid properties used in all simulations, which correspond to hydrogen ganglia in de-ionized water at 40\textdegree C. In iPNM, open boundaries are modeled by holding the wetting-phase pressure at $p_w\!=\!1$~atm, allowing displaced water to leave or enter the domain.
Finally, iPNM throats connected to open boundaries are shrunk by a factor of $10^3$ per cross-sectional dimension to prevent ganglia from entering or exiting the domain. The impact of this shrinking on GNM is felt through $C_\partial$, which is the sum of all boundary-throat conductivities.

\begin{table}[t!]
\centering
\caption{Properties of the porous media (Fig.~\ref{fig:validate}) and fluids used in all simulations. The latter corresponds to hydrogen in de-ionized water at 40\textdegree C.}
\label{tab:setup}
\setlength{\tabcolsep}{6pt}
\renewcommand{\arraystretch}{1.2}
{\small
\begin{tabular}{llccccl}
\hline
Property              & Symbol     & D2D                 & B2D                 & S2D                 & S3D & Unit \\
\hline
Image resolution      & $\Delta x$ & $7.99\times10^{-5}$ & $8.76\times10^{-5}$ & $1.24\times10^{-4}$ & $4.00\times10^{-4}$ & cm \\
Gap thickness         & $g$        & $1.52\times10^{-3}$ & $1.05\times10^{-3}$ & $7.80\times10^{-4}$ & -- & cm \\
Image size            & --         & $1499^2$ & $1160^2$ & $3421^2$ & $300^3$ & vox \\
Domain size           & --         & $0.120\times0.120$  & $0.102\times0.102$  & $0.425\times0.425$  & $0.120^3$ & cm \\
Porosity              & $\phi$     & $0.486$ & $0.315$ & $0.307$ & $0.211$ & -- \\
GNM nodes             & --         & 1,420 & 775 & 6,721 & 1,466 & -- \\
GNM links             & --         & 1,721 & 950 & 8,487 & 2,061 & -- \\
iPNM pores            & --         & 354 & 144 & 1,068 & 340 & -- \\
iPNM throats          & --         & 652 & 188 & 1,880 & 837 & -- \\
Initial ganglia       & $n_G$        & 74 & 22 & 159 & 53 & -- \\
\hline
Temperature           & $T$        & \multicolumn{4}{c}{$313.15$}          & K \\
Surface tension       & $\sigma$   & \multicolumn{4}{c}{$68.9$}            & dyn/cm \\
Diffusion coefficient & $D_m$      & \multicolumn{4}{c}{$7.34\times10^{-5}$} & cm$^2$/s \\
Henry's constant      & $H$        & \multicolumn{4}{c}{$7.51\times10^{10}$} & dyn/cm$^2$ \\
Vapor pressure        & $p_v$      & \multicolumn{4}{c}{$7.36\times10^{4}$}  & dyn/cm$^2$ \\
Water molar density   & $\rho_w$   & \multicolumn{4}{c}{$0.055$}           & mol/cm$^3$ \\
H$_2$ molar density   & $\rho_b$   & \multicolumn{4}{c}{$3.89\times10^{-5}$} & mol/cm$^3$ \\
\hline
\end{tabular}}
\end{table}

\section{Results}
\label{sec:results}

We compare GNM against iPNM in the four porous domains shown in Fig.~\ref{fig:validate} under the three scenarios of Ostwald ripening, dissolution, and growth described in Section~\ref{sec:valid}. Four population-level quantities are computed and plotted versus time: the population-mean curvature $\bar{\kappa}$ (Eq.~\ref{eq:nostor2}), the number of ganglia $n_G$, the total non-wetting-phase volume $V_t\!=\!\sum_i V_{b,i}$, and the total moles $N_t$ (Eq.~\ref{eq:meanfield}) consisting of the moles dissolved in the water and those in the ganglia. In GNM a tethered set counts as one ganglion, so $n_G$ is comparable to iPNM's count of connected non-wetting regions.

\begin{table}[t!]
\centering
\caption{Summary of initial ($t_i$), intermediate ($t_m$), and late ($t_f$) times, at which snapshots are shown in subsequent figures. For each domain and scenario, we also include at $t_f$ the ganglion count $n_G$, mean curvature $\bar{\kappa}$, and total moles relative to the initial value $N_t/N_t^0$, from GNM and iPNM.}
\label{tab:results}
\setlength{\tabcolsep}{6pt}
\renewcommand{\arraystretch}{1.2}
{\small
\begin{tabular}{llccccccccc}
\hline
Scenario & Domain & $t_i$ (s) & $t_m$ (s) & $t_f$ (s) & \multicolumn{2}{c}{$n_G$} & \multicolumn{2}{c}{$\bar{\kappa}$ (cm$^{-1}$)} & \multicolumn{2}{c}{$N_t/N_t^0$} \\
 & & & & & GNM & iPNM & GNM & iPNM & GNM & iPNM \\
\hline
Ripening    & B2D & $10^{-4}$ & $1.3\times10^{3}$ & $1.1\times10^{5}$ & 12 & 13 & 2449 & 2447 & 1.000 & 1.000 \\
            & D2D & $10^{-4}$ & $8.7\times10^{2}$ & $9.3\times10^{4}$ & 45 & 44 & 1892 & 1949 & 1.000 & 1.000 \\
            & S2D & $10^{-4}$ & $3.6\times10^{3}$ & $1.0\times10^{6}$ & 102 & 98 & 2950 & 3028 & 1.000 & 1.000 \\
            & S3D & $10^{-4}$ & $8.6\times10^{3}$ & $1.0\times10^{7}$ & 12 & 13 & 380 & 441 & 1.000 & 1.001 \\
\hline
Dissolution & B2D & $10^{-4}$ & $6.2\times10^{3}$ & $1.5\times10^{4}$ & 2 & 2 & 2141 & 2103 & 0.341 & 0.343 \\
            & D2D & $10^{-4}$ & $6.8\times10^{3}$ & $2.6\times10^{4}$ & 7 & 5 & 2098 & 1782 & 0.172 & 0.168 \\
            & S2D & $10^{-4}$ & $5.6\times10^{4}$ & $1.2\times10^{5}$ & 5 & 19 & 3425 & 2829 & 0.069 & 0.186 \\
            & S3D & $10^{-4}$ & $1.2\times10^{4}$ & $9.0\times10^{6}$ & 5 & 6 & 292 & 454 & 0.525 & 0.568 \\
\hline
Growth      & B2D & $10^{-4}$ & $7.9\times10^{2}$ & $5.0\times10^{3}$ & 7 & 2 & 3600 & 3299 & 2.30 & 2.32 \\
            & D2D & $10^{-4}$ & $7.9\times10^{2}$ & $5.0\times10^{3}$ & 25 & 26 & 2365 & 2352 & 1.91 & 1.90 \\
            & S2D & $10^{-4}$ & $2.2\times10^{3}$ & $1.4\times10^{4}$ & 119 & 73 & 4703 & 3777 & 1.62 & 1.61 \\
			& S3D & $10^{-4}$ & $1.3\times10^{5}$ & $3.0\times10^{6}$ & 29 & 25 & 586 & 634 & 1.70 & 1.62 \\
\hline
\end{tabular}}
\end{table}

\subsection{Ostwald ripening}
\label{sec:res_ripen}

The ripening scenario primarily validates the mean-field approximation of the wetting phase, the assumption that the dissolved concentration $X_m$ is uniform throughout the water. iPNM serves as the reference because it resolves the concentration field explicitly, and thereby the mass transfer between each ganglion and the surrounding water.
We initialize simulations by setting the water concentration to $X_m\!=\!X_{mo}$ and imposing a closed boundary ($C_\partial\!=\!0$ in GNM, no-flux BCs in iPNM). The latter ensures that mass transfer occurs only between ganglia, from those with high interfacial curvature to those with low interfacial curvature, and not between ganglia and the boundary.

\begin{figure}[t!]
\centering
\hspace*{-1.2cm}
\includegraphics[width=1.1\linewidth]{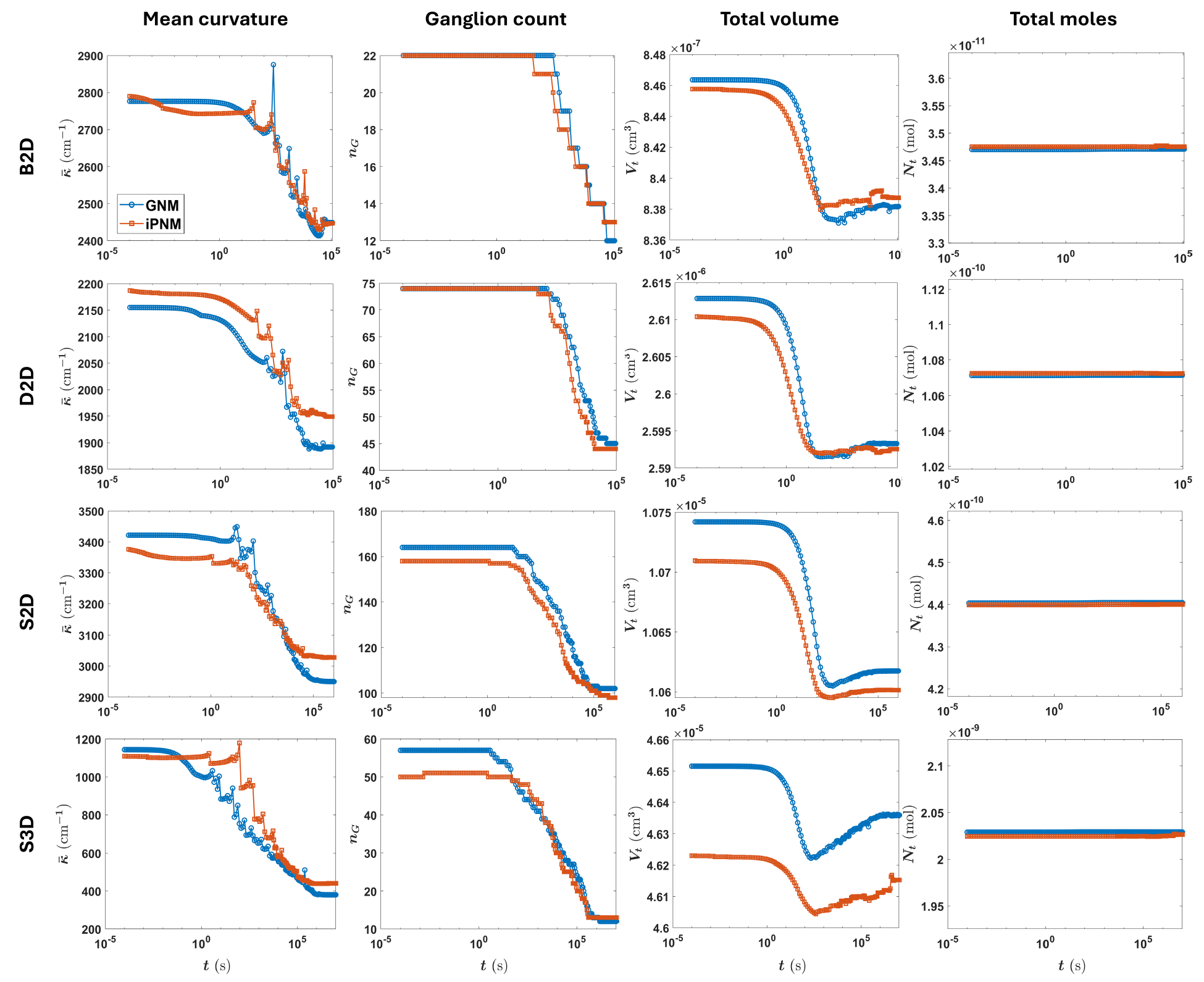}
\caption{Ostwald ripening scenario: mean curvature $\bar{\kappa}$, ganglion count $n_G$, total ganglion volume $V_t$, and total moles $N_t$ versus time in B2D, D2D, S2D, and S3D domains (rows), as predicted by GNM (blue circles) and iPNM (orange squares). Note the time axis is logarithmic.}
\label{fig:ripen_curves}
\end{figure}

\begin{figure}[t!]
\centering
\hspace*{-1.2cm}
\includegraphics[width=1.0\linewidth]{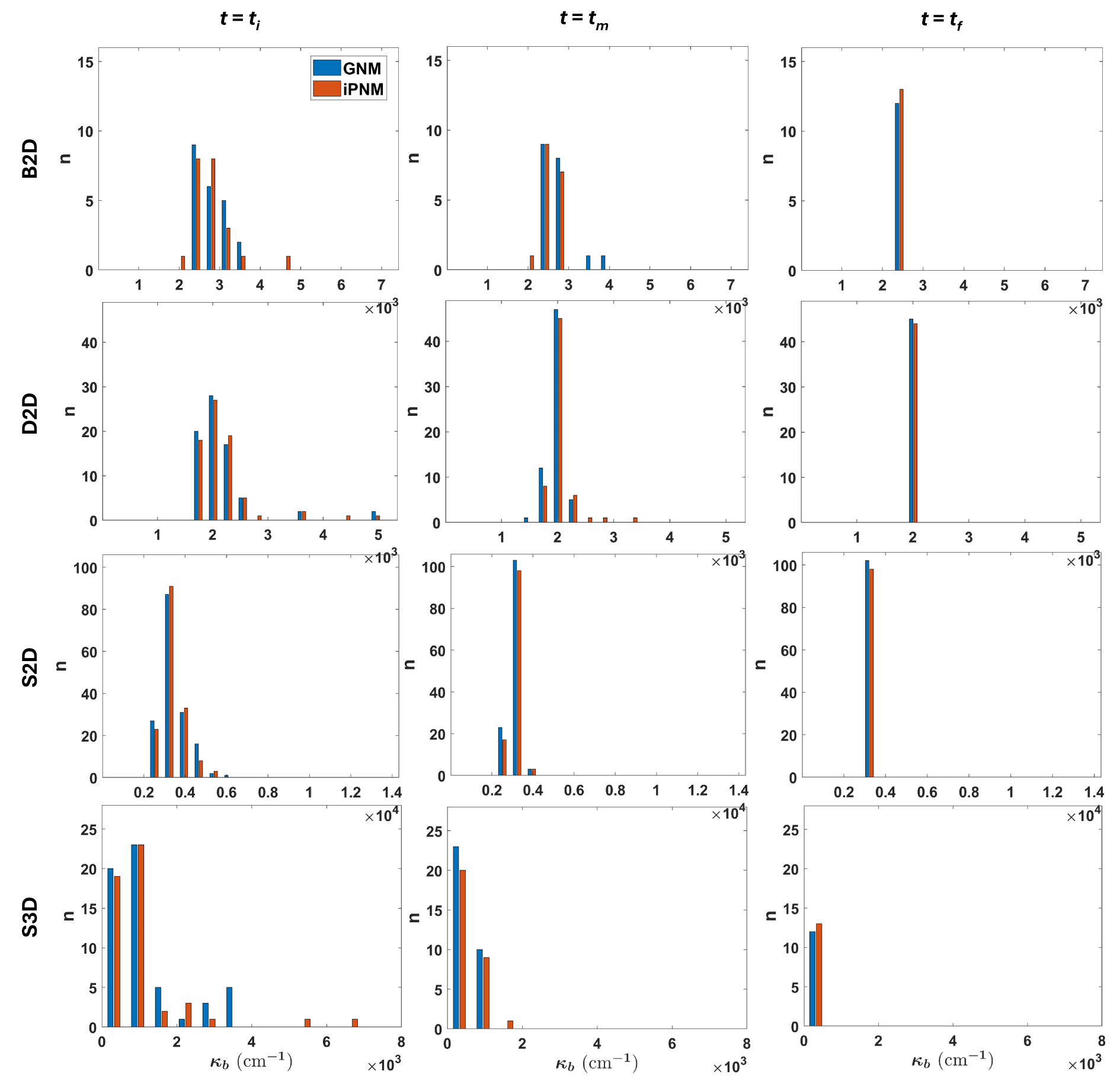}
\caption{Ostwald ripening scenario: number distributions of ganglion curvature $\kappa_b$ at $t\!=\!t_i$, $t\!=\!t_m$, and $t\!=\!t_f$ (columns) in B2D, D2D, S2D, and S3D domains (rows), as predicted by GNM (blue) and iPNM (orange); $n$ is the number of ganglia per bin. Snapshot times are listed in Table~\ref{tab:results}.}
\label{fig:ripen_dist_kb}
\end{figure}

\begin{figure}[t!]
\centering
\hspace*{-1.2cm}
\includegraphics[width=1.0\linewidth]{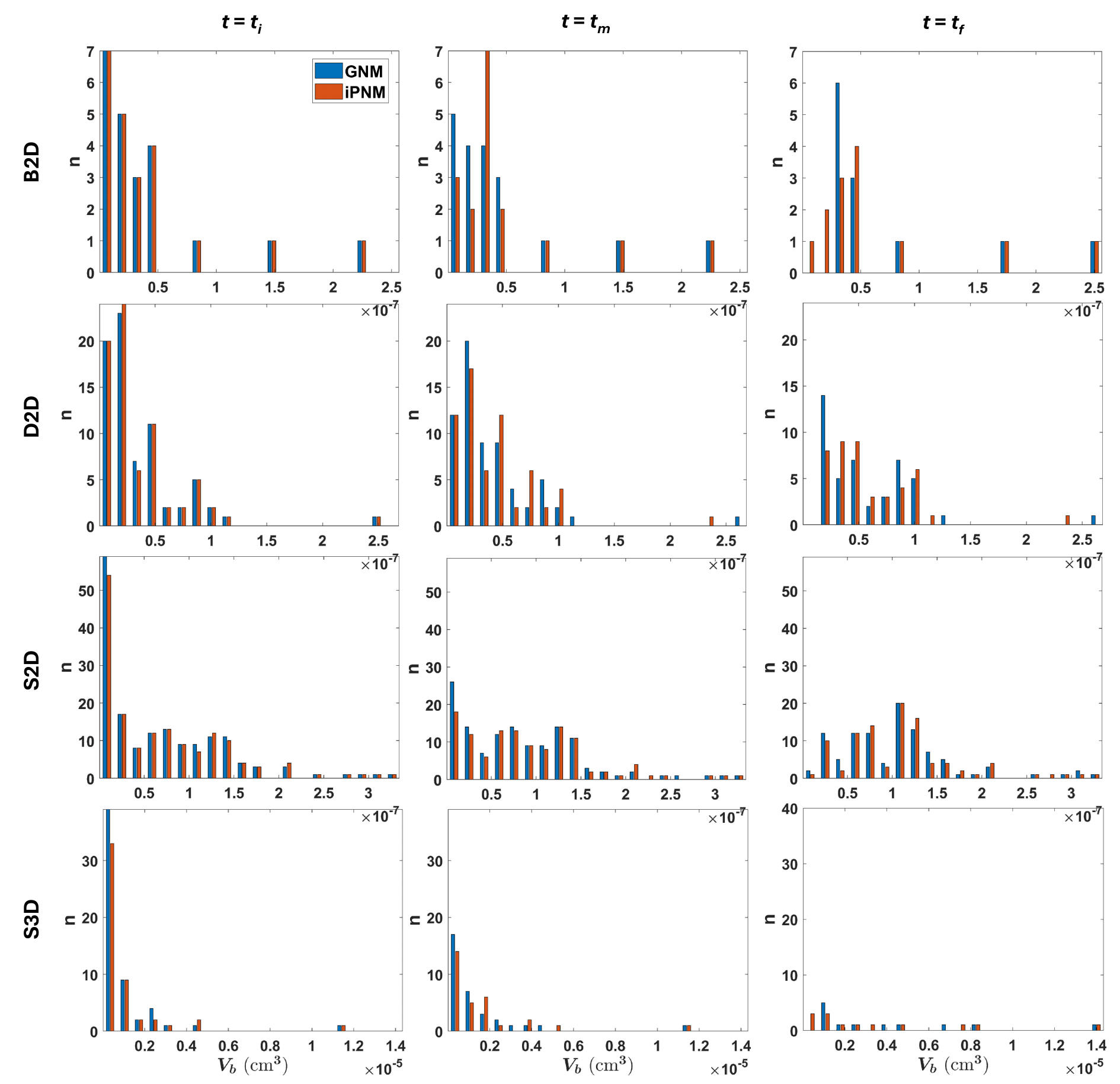}
\caption{Ostwald ripening scenario: number distributions of ganglion volume $V_b$ at $t\!=\!t_i$, $t\!=\!t_m$, and $t\!=\!t_f$ (columns) in B2D, D2D, S2D, and S3D domains (rows), as predicted by GNM (blue) and iPNM (orange); $n$ is the number of ganglia per bin. Snapshot times are listed in Table~\ref{tab:results}.}
\label{fig:ripen_dist_vb}
\end{figure}

\begin{figure}[p]
\centering
\includegraphics[height=0.95\textheight,keepaspectratio]{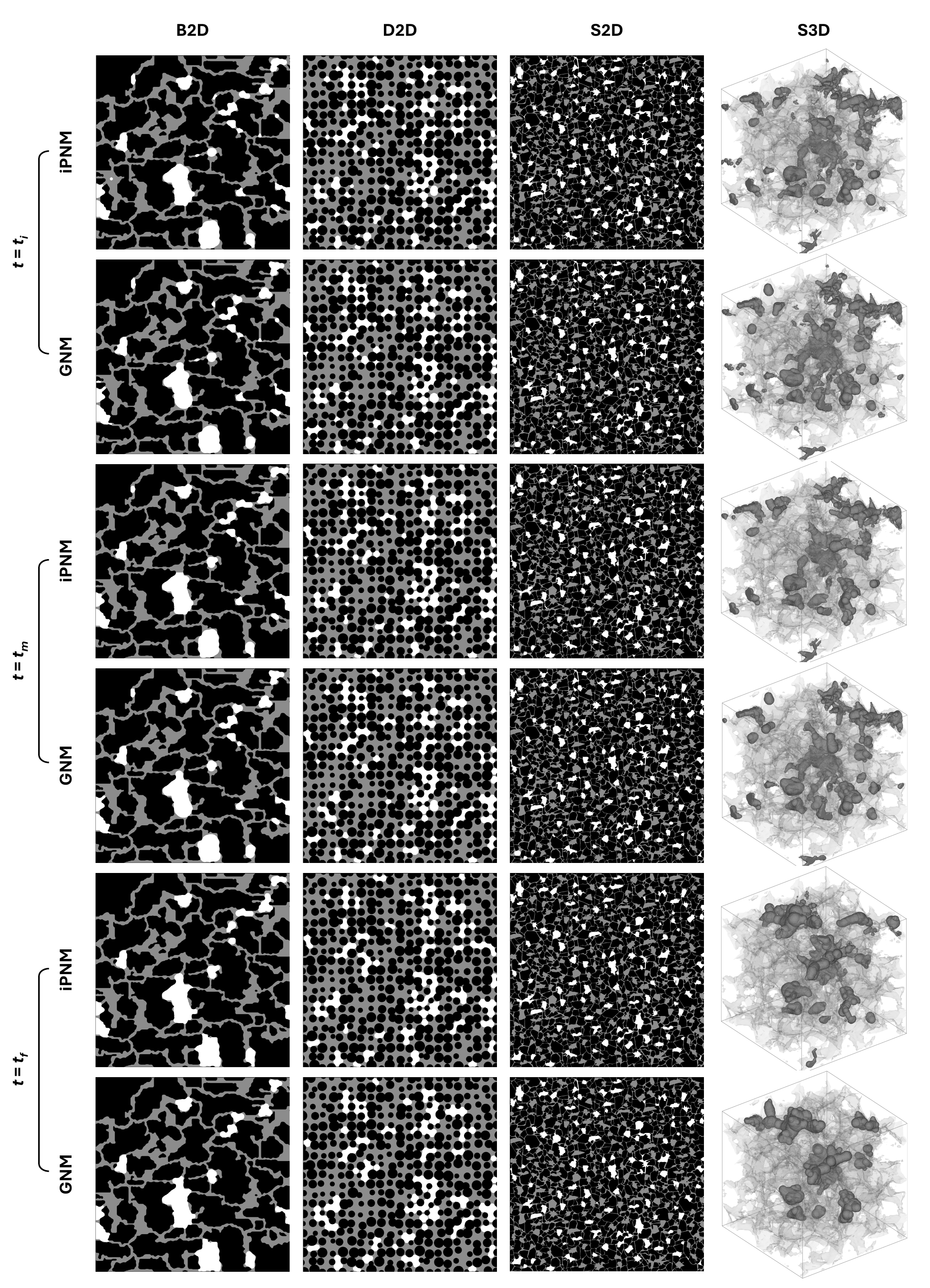}
\caption{Ostwald ripening scenario: spatial distribution of ganglia at $t\!=\!t_i$, $t\!=\!t_m$, and $t\!=\!t_f$ in the B2D, D2D, S2D, and S3D domains, as predicted by iPNM and GNM. In 2.5D, ganglia$=$white, water$=$gray, solid$=$black. In 3D, ganglia$=$gray, solid$=$translucent. Snapshot times are in Table~\ref{tab:results}.}
\label{fig:ripen_snaps}
\end{figure}

Fig.~\ref{fig:ripen_curves} depicts $\bar{\kappa}$, $n_G$, $V_t$ and $N_t$ versus time for the four domains, one row per domain, comparing the predictions of GNM to those of iPNM. Figs.~\ref{fig:ripen_dist_kb}--\ref{fig:ripen_dist_vb} show the number distributions of ganglion curvature and ganglion volume over the population at the initial time ($t\!=\!t_i$), an intermediate time ($t\!=\!t_m$), and a late time ($t\!=\!t_f$) at which the population has reached equilibrium and $\bar{\kappa}$ has plateaued.
These times are listed in Table~\ref{tab:results} for each case, where $t_i\!=\!10^{-4}$ corresponds to the first time step taken by both models in all cases.
Fig.~\ref{fig:ripen_snaps} shows the spatial distribution of the ganglia predicted by both models at the same three snapshots, GNM drawn with the enhanced rendering discussed in Section~\ref{sec:visualize}.

GNM and iPNM agree well across all four domains. In each domain, $\bar{\kappa}$ decreases and asymptotes to a constant, as it must, since the surviving ganglia share one curvature at equilibrium. The $n_G$ from the two models stay within a few ganglia of each other, ending at 12 for GNM versus 13 for iPNM in B2D, 45 versus 44 in D2D, 102 versus 98 in S2D and 12 versus 13 in S3D at $t_f$. The $\bar{\kappa}$ predictions at $t_f$ agree to within 3\% in the 2.5D domains and 14\% in S3D. $N_t$ is constant in both models, since the system is closed and ripening only exchanges mass among ganglia. By contrast, $V_t$ decreases slightly at first and then plateaus, because the water starts in equilibrium with a flat interface ($X_m\!=\!X_{mo}$) and every curved ganglion gives up a small fraction of its volume to saturate it.
Notice the $V_t$ in iPNM sits slightly below that of GNM at early times. This is because at $t\!=\!t_i$, iPNM imposes local equilibrium between each ganglion and the water in every pore~\cite{laku2026ipnm}, dissolving a small volume at once. The same offset appears later in Section~\ref{sec:res_diss} for the dissolution scenario, and with opposite sign in Section~\ref{sec:res_grow} for the growth scenario due to the water being supersaturated.
Figs.~\ref{fig:ripen_dist_kb}--\ref{fig:ripen_dist_vb} show that the curvature distributions converge towards a Dirac delta and the volume distributions coarsen slightly, as the smallest ganglia---most confined to one pore---dissolve. The spatial distributions in Fig.~\ref{fig:ripen_snaps} are consistent with this observation. Table~\ref{tab:results} lists $n_G$, $\bar{\kappa}$, and $N_t/N_t^0$ at $t_f$ for every case in GNM and iPNM.

\subsection{Dissolution}
\label{sec:res_diss}

The dissolution scenario introduces the exchange of solute between the water and the boundary, and it primarily validates the undershoot rule of GNM. As ganglia shrink, they move down the tree graph (Fig.~\ref{fig:coarse}) and fragment as they pass through junction nodes. The shrinking ganglia also continue to exchange mass with one another through the mean field. We use iPNM as the reference because it explicitly resolves the mass transfer between ganglia and the boundary, as well as among the ganglia themselves. We initialize the water at $X_m\!=\!X_{mo}/10$ and open the boundary at the same concentration ($X_\partial\!=\!X_{mo}/10$ in GNM, Dirichlet BCs in iPNM), so the water is undersaturated with respect to every interface. Therefore, all ganglia dissolve from the start, and the solute they release leaves through the boundary.

\begin{figure}[t!]
\centering
\hspace*{-1.2cm}
\includegraphics[width=1.1\linewidth]{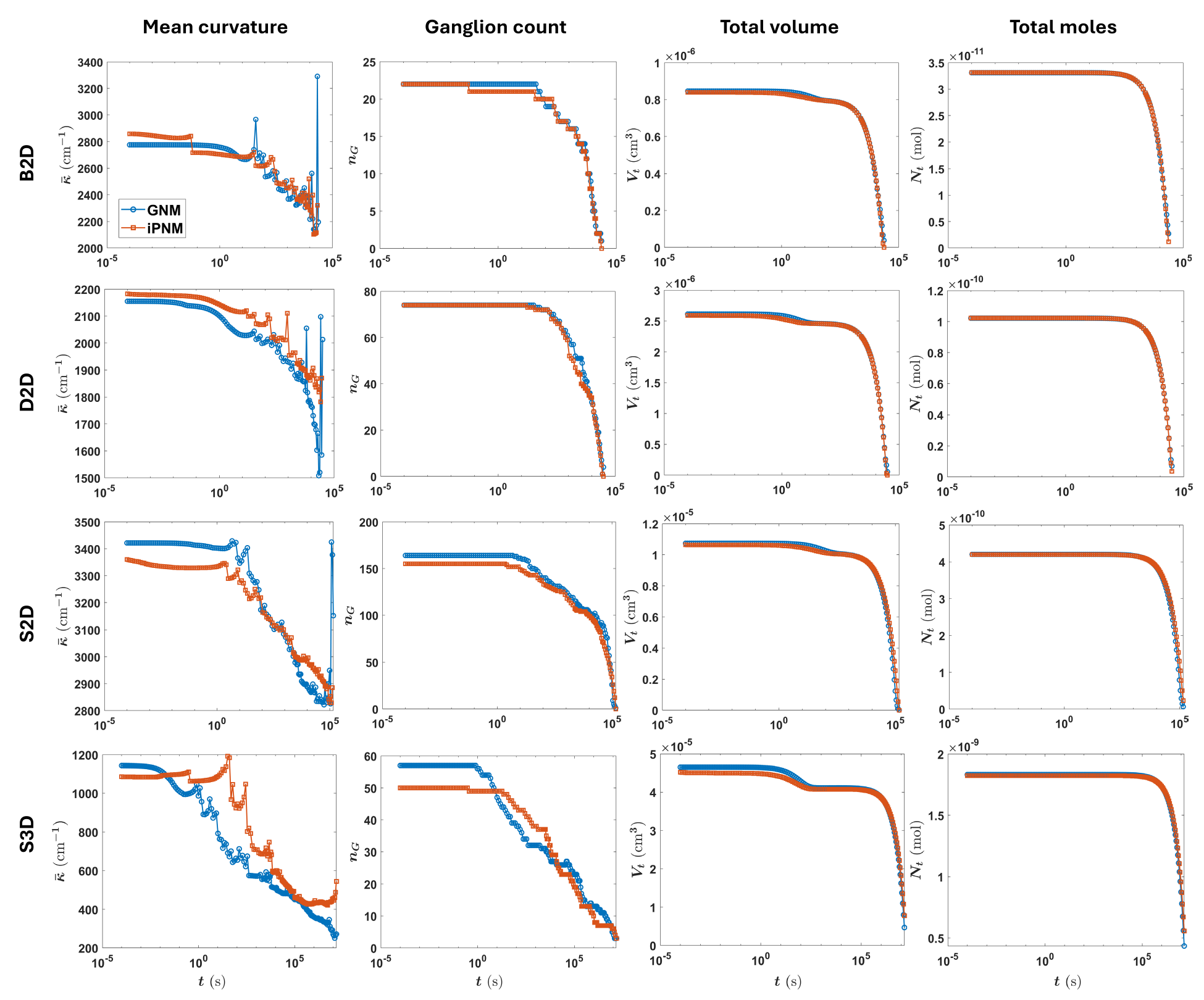}
\caption{Dissolution scenario: mean curvature $\bar{\kappa}$, ganglion count $n_G$, total ganglion volume $V_t$, and total moles $N_t$ versus time in B2D, D2D, S2D, and S3D domains (rows), as predicted by GNM (blue circles) and iPNM (orange squares). Note the time axis is logarithmic.}
\label{fig:diss_curves}
\end{figure}

\begin{figure}[t!]
\centering
\hspace*{-1.2cm}
\includegraphics[width=1.0\linewidth]{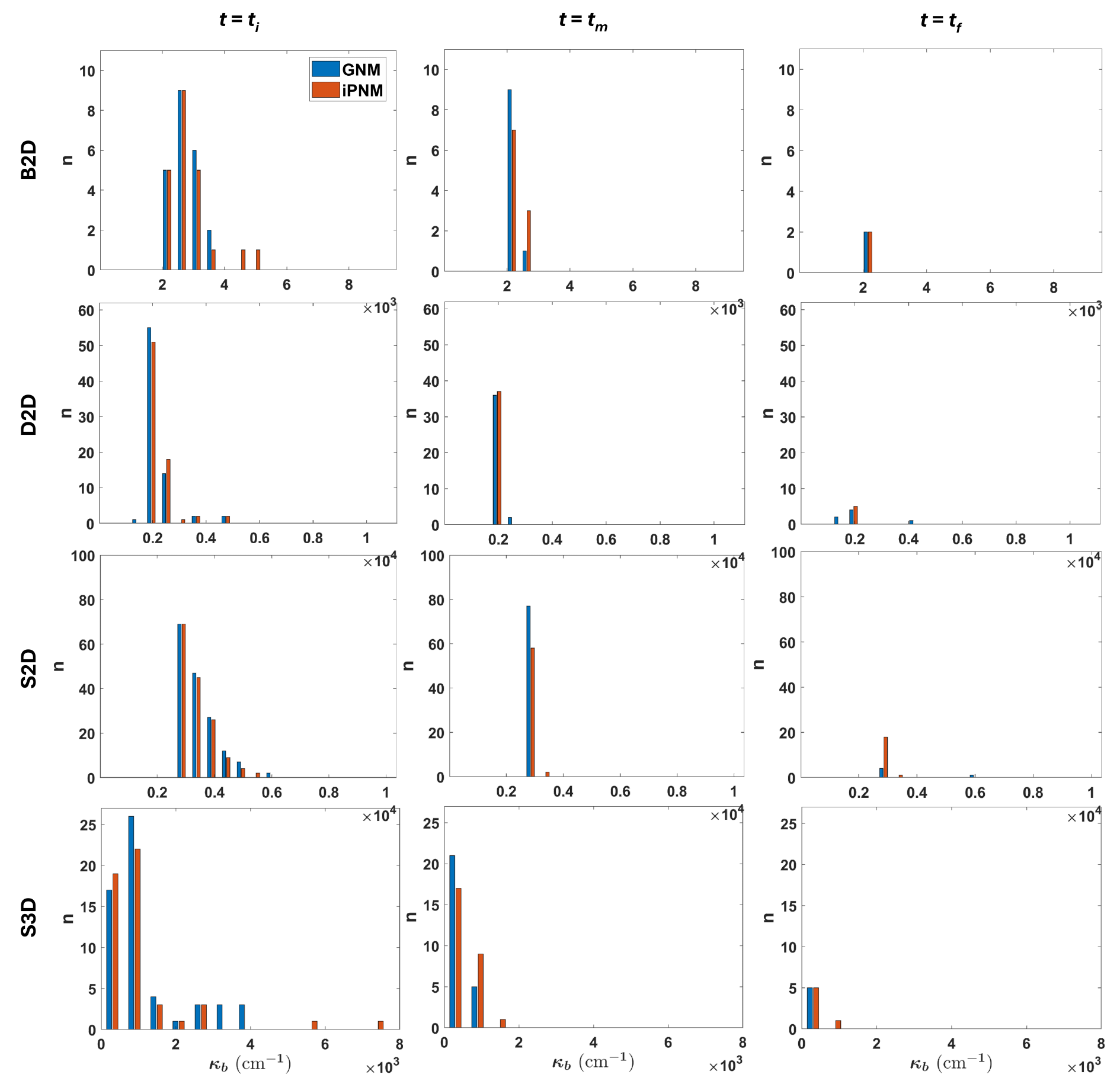}
\caption{Dissolution scenario: number distributions of ganglion curvature $\kappa_b$ at $t\!=\!t_i$, $t\!=\!t_m$, and $t\!=\!t_f$ (columns) in B2D, D2D, S2D, and S3D domains (rows), as predicted by GNM (blue) and iPNM (orange); $n$ is the number of ganglia per bin. Snapshot times are listed in Table~\ref{tab:results}.}
\label{fig:diss_dist_kb}
\end{figure}

\begin{figure}[t!]
\centering
\hspace*{-1.2cm}
\includegraphics[width=1.0\linewidth]{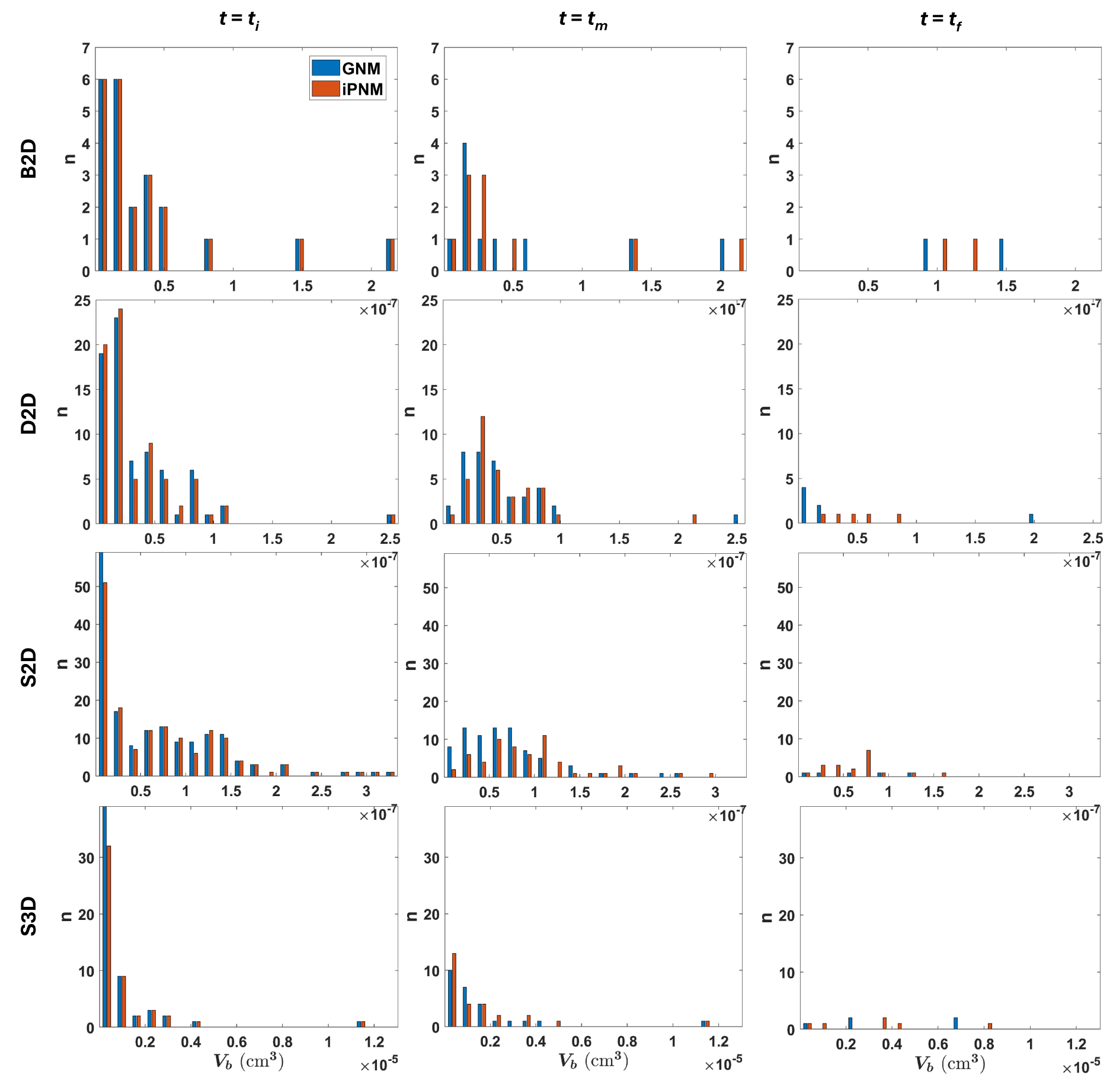}
\caption{Dissolution scenario: number distributions of ganglion volume $V_b$ at $t\!=\!t_i$, $t\!=\!t_m$, and $t\!=\!t_f$ (columns) in B2D, D2D, S2D, and S3D domains (rows), as predicted by GNM (blue) and iPNM (orange); $n$ is the number of ganglia per bin. Snapshot times are listed in Table~\ref{tab:results}.}
\label{fig:diss_dist_vb}
\end{figure}

\begin{figure}[p]
\centering
\includegraphics[height=0.95\textheight,keepaspectratio]{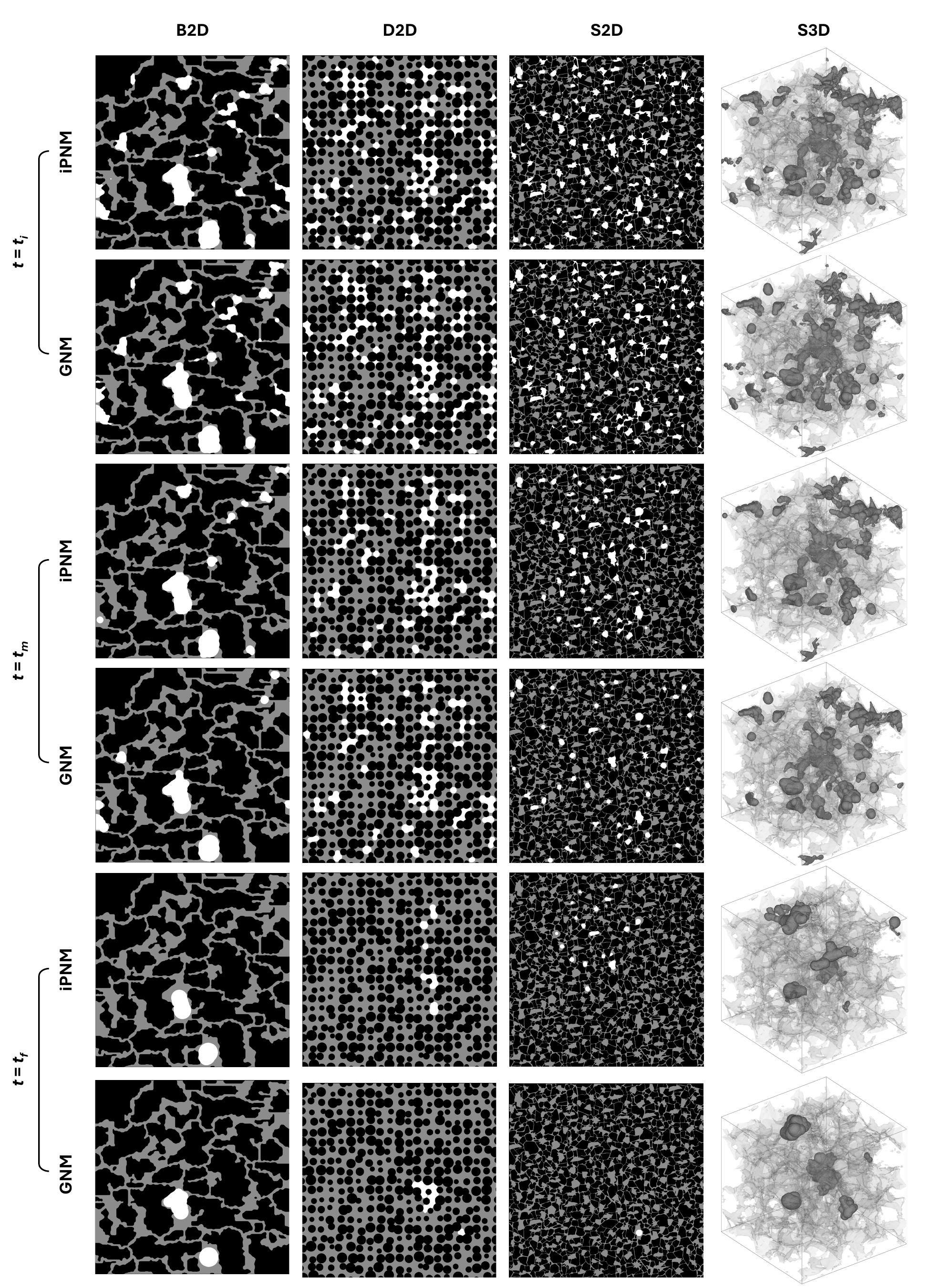}
\caption{Dissolution scenario: spatial distribution of ganglia at $t\!=\!t_i$, $t\!=\!t_m$, and $t\!=\!t_f$ in the B2D, D2D, S2D, and S3D domains, as predicted by iPNM and GNM. In 2.5D, ganglia$=$white, water$=$gray, solid$=$black. In 3D, ganglia$=$gray, solid$=$translucent. Snapshot times are in Table~\ref{tab:results}.}
\label{fig:diss_snaps}
\end{figure}

As in Section~\ref{sec:res_ripen}, Fig.~\ref{fig:diss_curves} depicts $\bar{\kappa}$, $n_G$, $V_t$ and $N_t$ versus time for the four domains. Figs.~\ref{fig:diss_dist_kb}--\ref{fig:diss_dist_vb} show the number distributions of ganglion curvature and ganglion volume, and Fig.~\ref{fig:diss_snaps} illustrates the spatial distributions of ganglia at $t\!=\!t_i$, $t\!=\!t_m$ and $t\!=\!t_f$.
Here, $t_f$ is taken shortly after the $n_G$ in GNM falls below a tenth of its initial value (Table~\ref{tab:results}).

GNM and iPNM agree well across all four domains. The ganglion count $n_G$ decreases in both models until the population vanishes.
By $t_f$, iPNM is down to its last 2, 5, 19, and 6 ganglia in B2D, D2D, S2D, and S3D, and GNM to 2, 7, 5, and 5.
The total non-wetting-phase volume $V_t$ and the total moles $N_t$ fall together in both models, since the solute released by the ganglia leaves through the boundary rather than accumulating in water.
Table~\ref{tab:results} shows that by $t_f$ between 7 and 57\% of the initial moles remain.
The mean curvature $\bar{\kappa}$ decreases over most of the run, because the smallest and most curved ganglia vanish first, and it spikes only near $t_f$ as the last surviving ganglion shrinks to a point.
Apart from the early-time offset noted in Section~\ref{sec:res_ripen}, $V_t$ and $N_t$ of the two models agree to within 2\% and 1\% until the final decade of the run, when few ganglia remain.
Volume and curvature distributions in Figs.~\ref{fig:diss_dist_kb}--\ref{fig:diss_dist_vb} show small high-curvature ganglia dissolve first, followed by large low-curvature ganglia until population collapse. The spatial plots in Fig.~\ref{fig:diss_snaps} depict the same process.
Table~\ref{tab:results} lists $n_G$, $\bar{\kappa}$, and $N_t/N_t^0$ at $t_f$ for every case in GNM and iPNM.

\subsection{Growth}
\label{sec:res_grow}

The growth scenario primarily validates the overshoot rule of GNM, the most elaborate of the three in Section~\ref{sec:rules}. As ganglia grow, they move up the tree graph (Fig.~\ref{fig:coarse}), and at junction nodes, the spill, seed, fire, tether, and snap-off rules of Section~\ref{sec:rules} decide how they invade neighboring pores and merge or fragment in the process. Growing ganglia also continue to exchange mass with one another through the mean field. We use iPNM as the reference because it resolves the invasion of individual pores and throats explicitly, as well as the transfer of mass between ganglia and the boundary and among the ganglia themselves. We initialize the water at $X_m\!=\!10\,X_{mo}$ and open the boundary at the same concentration ($X_\partial\!=\!10\,X_{mo}$ in GNM, Dirichlet BCs in iPNM), so the water is supersaturated with respect to every interface. Therefore, all ganglia grow from the start, and the solute they absorb enters through the boundary.

\begin{figure}[t!]
\centering
\hspace*{-1.2cm}
\includegraphics[width=1.1\linewidth]{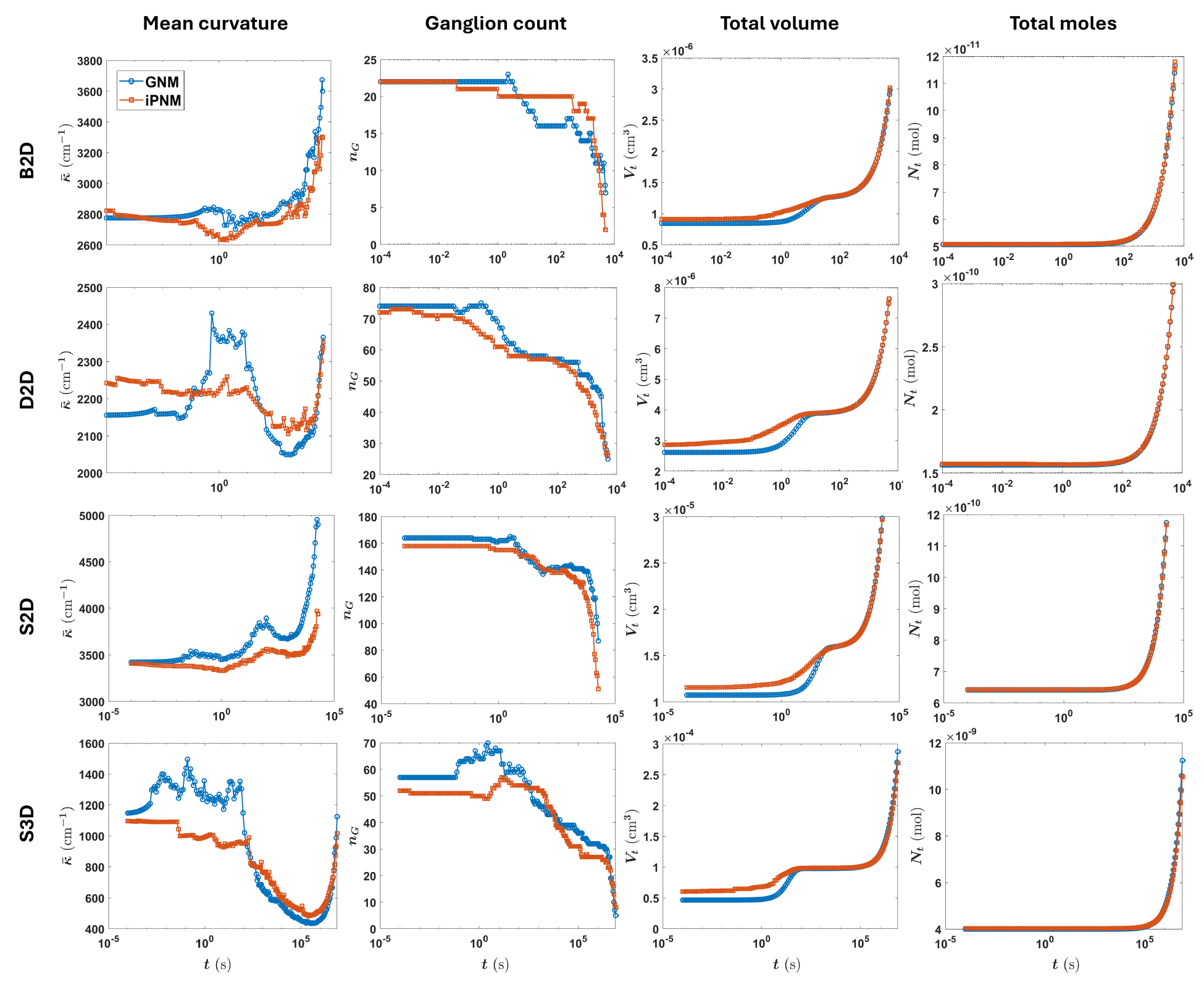}
\caption{Growth scenario: mean curvature $\bar{\kappa}$, ganglion count $n_G$, total ganglion volume $V_t$, and total moles $N_t$ versus time in B2D, D2D, S2D, and S3D domains (rows), as predicted by GNM (blue circles) and iPNM (orange squares). Note the time axis is logarithmic.}
\label{fig:grow_curves}
\end{figure}

\begin{figure}[t!]
\centering
\hspace*{-1.2cm}
\includegraphics[width=1.0\linewidth]{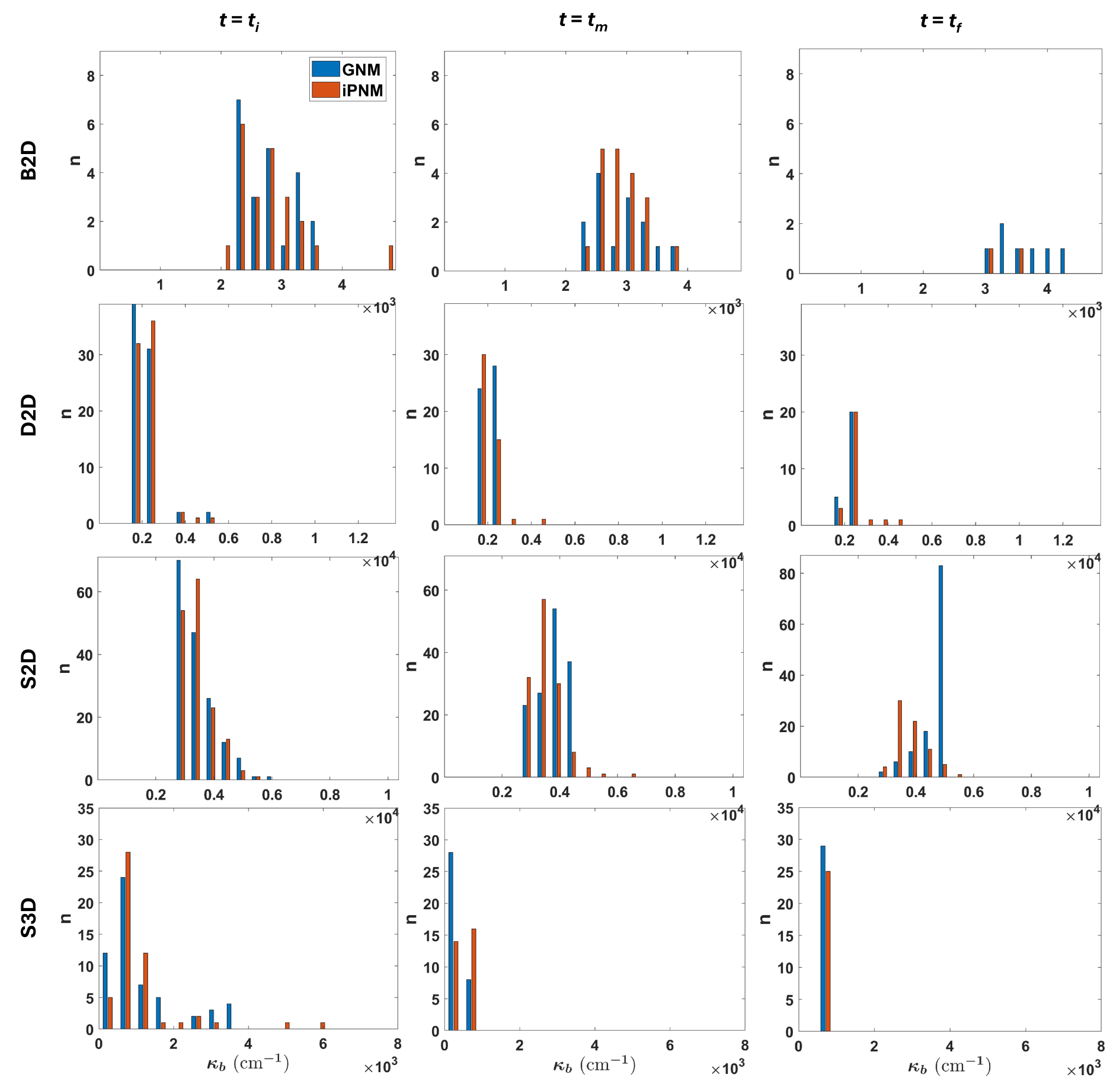}
\caption{Growth scenario: number distributions of ganglion curvature $\kappa_b$ at $t\!=\!t_i$, $t\!=\!t_m$, and $t\!=\!t_f$ (columns) in B2D, D2D, S2D, and S3D domains (rows), as predicted by GNM (blue) and iPNM (orange); $n$ is the number of ganglia per bin. Snapshot times are listed in Table~\ref{tab:results}.}
\label{fig:grow_dist_kb}
\end{figure}

\begin{figure}[t!]
\centering
\hspace*{-1.2cm}
\includegraphics[width=1.0\linewidth]{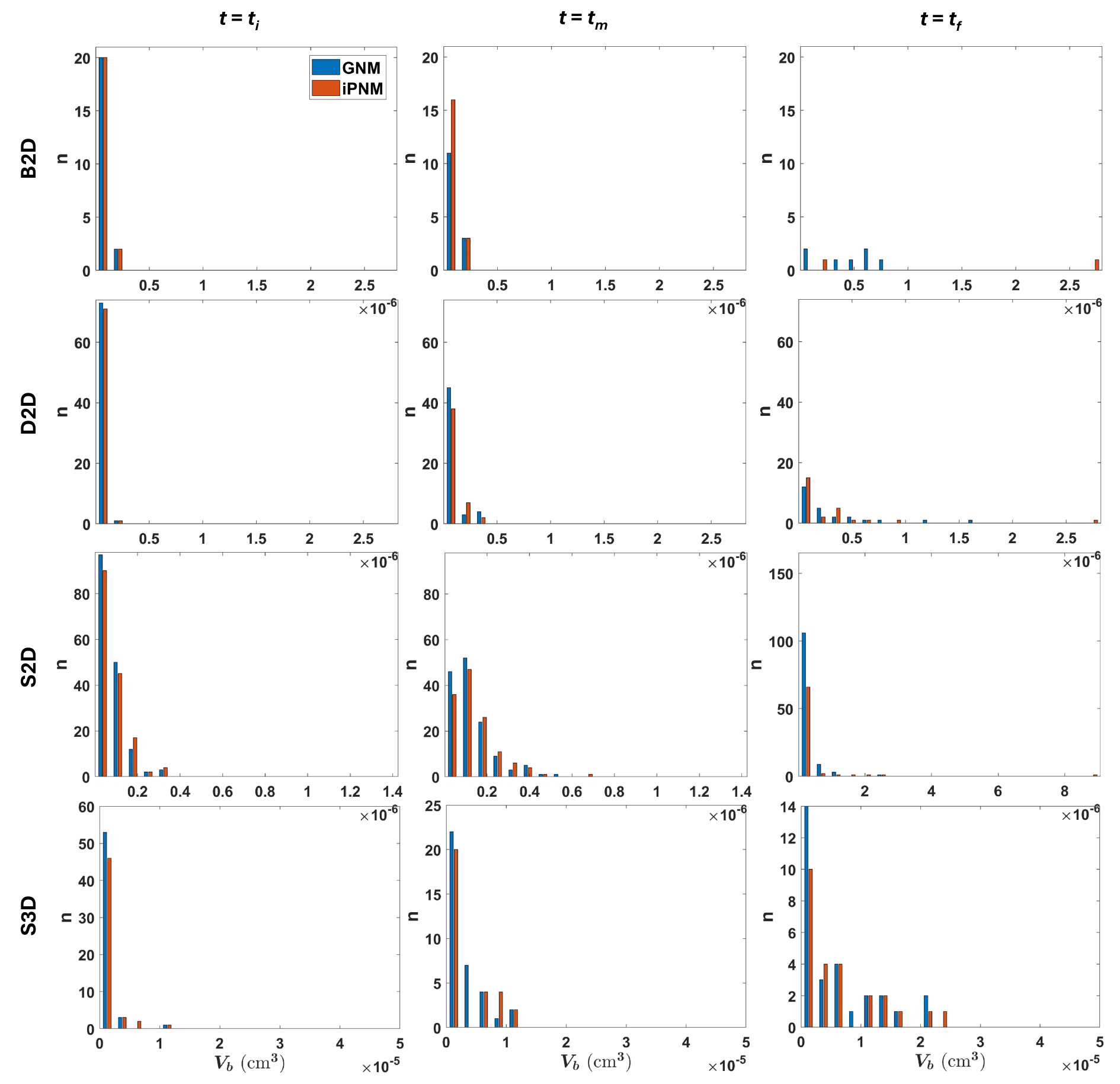}
\caption{Growth scenario: number distributions of ganglion volume $V_b$ at $t\!=\!t_i$, $t\!=\!t_m$, and $t\!=\!t_f$ (columns) in B2D, D2D, S2D, and S3D domains (rows), as predicted by GNM (blue) and iPNM (orange); $n$ is the number of ganglia per bin. Snapshot times are listed in Table~\ref{tab:results}.}
\label{fig:grow_dist_vb}
\end{figure}

\begin{figure}[p]
\centering
\includegraphics[height=0.95\textheight,keepaspectratio]{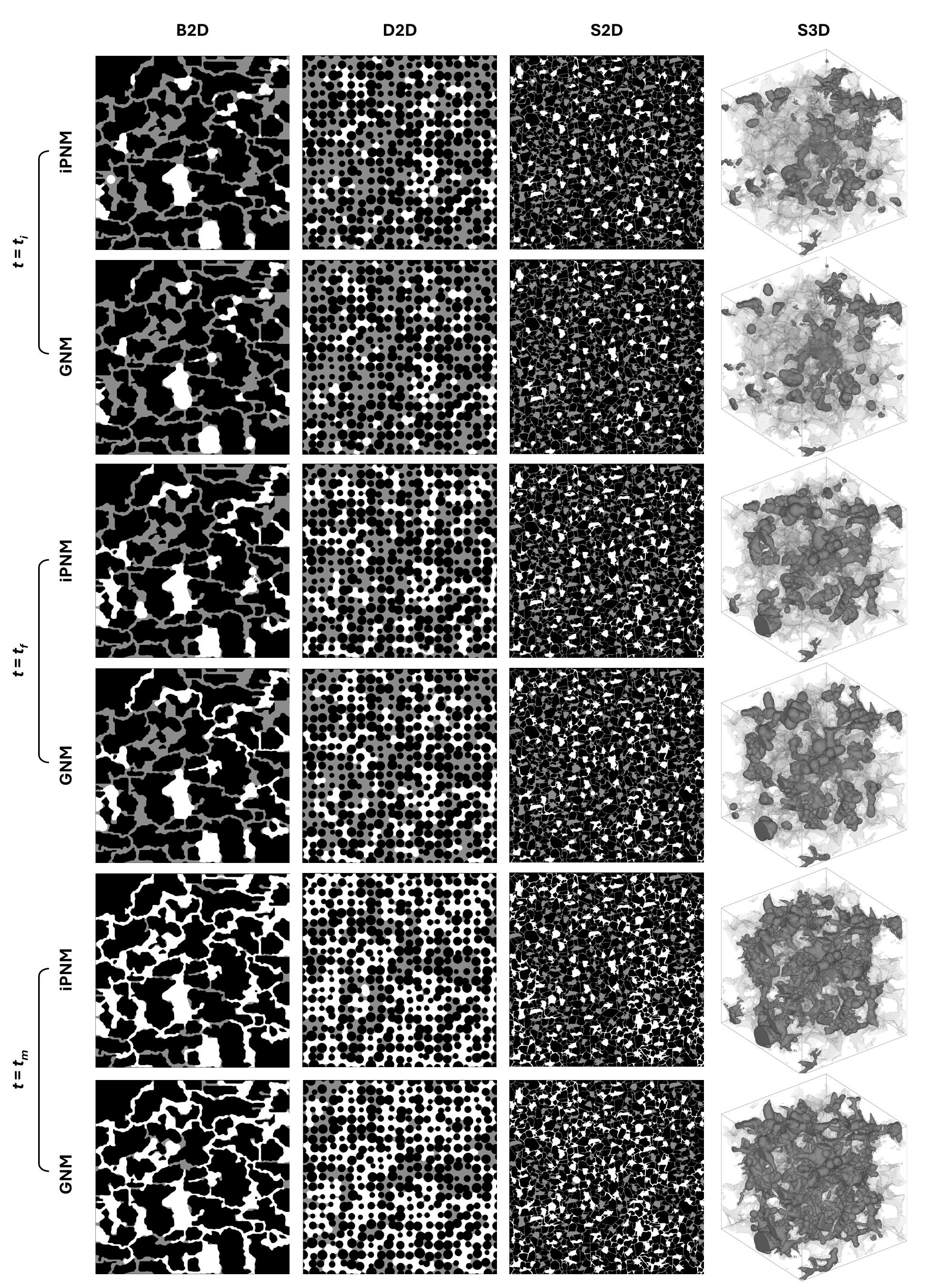}
\caption{Growth scenario: spatial distribution of ganglia at $t\!=\!t_i$, $t\!=\!t_m$, and $t\!=\!t_f$ in the B2D, D2D, S2D, and S3D domains, as predicted by iPNM and GNM. In 2.5D, ganglia$=$white, water$=$gray, solid$=$black. In 3D, ganglia$=$gray, solid$=$translucent. Snapshot times are in Table~\ref{tab:results}.}
\label{fig:grow_snaps}
\end{figure}

As in Section~\ref{sec:res_ripen}, Fig.~\ref{fig:grow_curves} depicts $\bar{\kappa}$, $n_G$, $V_t$ and $N_t$ versus time for the four domains. Figs.~\ref{fig:grow_dist_kb}--\ref{fig:grow_dist_vb} show the number distributions of ganglion curvature and ganglion volume, and Fig.~\ref{fig:grow_snaps} illustrates the spatial distributions of ganglia at $t\!=\!t_i$, $t\!=\!t_m$ and $t\!=\!t_f$ (Table~\ref{tab:results}).
The simulations are run until the non-wetting-phase saturation reaches 0.87, 0.72, 0.61, and 0.47 in B2D, D2D, S2D, and S3D, respectively, which defines the final time $t_f$ in Table~\ref{tab:results}.

GNM and iPNM agree well across all four domains. The ganglion count $n_G$ decreases in both models as growing ganglia merge, ending at 7 for GNM versus 2 for iPNM in B2D, 25 versus 26 in D2D, 119 versus 73 in S2D, and 29 versus 25 in S3D at $t_f$.
Over the whole run, $n_G$ of the two models differs on average by 1.8, 3.6, 7.7, and 6.3 ganglia in B2D, D2D, S2D, and S3D (i.e., at most 12\% of the initial count).
The total non-wetting-phase volume $V_t$ and the total moles $N_t$ rise together in both models, as solute enters through the boundary and is absorbed by ganglia. Table~\ref{tab:results} shows $N_t/N_t^0$ at $t_f$ agrees to within 5\%.
The mean curvature $\bar{\kappa}$ is mostly flat except for a late-time rise, reflecting that ganglion growth is initially driven by invading new pores with roughly similar entry pressures, followed by filling the corners of already occupied pores.
The $\bar{\kappa}$ predictions differ by 3, 3, 5, and 17\% on average.
Notice the early-time offset in $V_t$, noted in Section~\ref{sec:res_ripen}, has the opposite sign here.
At $t\!=\!t_i$, iPNM equilibrates ganglia with the water in each pore and thereby inflates the initial volume by 8 to 10\% in the 2.5D domains and 30\% in S3D, whereas GNM grows $V_t$ gradually via the mean field. The two $V_t$ curves come within 5\% of each other by $t\!\approx\!50$~s.
Volume and curvature distributions in Figs.~\ref{fig:grow_dist_kb}--\ref{fig:grow_dist_vb} shift towards larger volumes and larger curvatures as ganglia grow and merge. The spatial plots in Fig.~\ref{fig:grow_snaps} depict the same process. Table~\ref{tab:results} summarizes $n_G$, $\bar{\kappa}$, and $N_t/N_t^0$ at $t_f$ for every case in both models.

\section{Discussion}
\label{sec:disc}

\subsection{Sources of discrepancy}
\label{sec:disc_discrep}

Taking iPNM as the reference, the deviations in GNM reflect two approximations. The first is the mean-field assumption, which asserts that the dissolved concentration is spatially uniform.
This is tested primarily by the ripening and dissolution scenarios. Under ripening, the boundary is closed and the assumption is reasonable, implied by the agreement in Section~\ref{sec:res_ripen}.
Under dissolution, the boundary is open to solute flux, causing a concentration gradient to develop between the domain's interior and its boundary, captured by iPNM. This leads to ganglia farthest from the boundary to dissolve last, as seen for S2D in Fig.~\ref{fig:diss_snaps}. The effect is negligible for other domains because they are smaller (i.e., fewer pores per dimension), making the formation of a depleted zone near the boundary less pronounced~\cite{mehmani2024deplete}.
Unlike iPNM, GNM does not capture such concentration gradients, resulting in ganglia that dissolve more uniformly even at late times. In a sense, GNM models regions of a large porous medium far away from external boundaries.

The second approximation pertains to the rules of motion on the ganglion network, postulated in Section~\ref{sec:rules}. The traversal and undershoot rules are simple and capture explicitly how a shrinking ganglion deforms and fragments within the pore space. The overshoot rule is far more complex, involving the selection of a spill path, a spill event, potential slide-down of the invading ganglion, formation and severing of tethers, and a possible cascade or fire. These are tested primarily by the growth scenario.
An overshoot is a reduced-order representation of a ganglion's invasion of a neighboring pore, which may be accompanied by the ganglion's retraction from another pore or its snap-off.
Figs.~\ref{fig:grow_curves}--\ref{fig:grow_snaps} show that the rule captures these complex dynamics following a pore invasion reasonably well.

Apart from the above two approximations, three additional sources of discrepancy between iPNM and GNM exist. First, iPNM resolves the internal pressure and velocity fields within each ganglion, whereas GNM assumes ganglia are in local capillary equilibrium (i.e., no internal flow and equal curvature of all interfaces). Second, in the first time step of iPNM, the non-wetting phase in each pore is forced to be in local thermodynamic equilibrium with the water in that pore, causing a small volume of ganglia to dissolve instantaneously. By contrast, mass transfer between ganglia and the mean field in GNM is kinetically controlled.
The result is the early-time offset in $V_t$ noted in Sections~\ref{sec:res_ripen}--\ref{sec:res_grow}, which disappears within tens of seconds.
Another contributor to this offset is the imperfect mapping of ganglia from the graph to the pores of iPNM at $t\!=\!0$ (Section~\ref{sec:valid}). Disconnected ganglia in GNM may touch on the image, which then map to a single connected body in iPNM.
In GNM, we tether such ganglia at $t\!=\!0$, but most tethers snap at the first step, raising $n_G$ above that of iPNM.
Third, while both iPNM and GNM avoid simplifying the pore geometry by using PMM to parameterize their inputs, iPNM applies it locally to each pore. The implication is that iPNM's derived curvature--saturation relations must be adjusted slightly to capture cases where an occupied pore has low non-wetting saturation \textit{and} is part of a multi-pore ganglion~\cite{laku2026ipnm}. In this respect, GNM is free from such post hoc adjustments.

\subsection{Computational complexity}
\label{sec:disc_cost}

Let us compare the computational complexity of the pore-network model to that of the ganglion network model.
iPNM advances two coupled nonlinear systems in a staggered fashion, one for flow and one for solute transport with ripening. Each is solved by Newton iterations over the pore network. The cost per time step is $O(n_p^{\theta}\, n_{newt})$, where $n_p$ is the number of pores, $n_{newt}$ the number of Newton iterations, and $\theta\!\in\!(1,3)$ depends on the linear solver. Hence, the cost of iPNM scales with the domain size, regardless of the number of ganglia (even if only a handful).
In contrast, GNM solves no system of equations.
Instead, each time step updates the state of ganglia explicitly on the tree graph, at a cost of $O(n_G)$. Thus, GNM's cost scales with the ganglion count $n_G$, not the number of pores or domain size.

In both models the cost grows linearly with the number of time steps, which is adaptive.
We note that in GNM, the time step formulated in Section~\ref{sec:numerics} is overly conservative, because it is limited by the shortest occupied link in the tree graph. This can result in many more time steps than necessary in trees that have a wide range of link lengths. We chose it here only to ensure fully converged results, but in practice, it can be increased many-fold without appreciably affecting the results. The reason is that the iterative sweeps of traversal, undershoot, and overshoot in Algorithm~\ref{alg:rules} carry a ganglion to its final link regardless of whether a time step moves the ganglion across one or multiple nodes.
Coarsening the tree at $f_\kappa\!=\!0.1$ then yields further speedup, whose magnitude is modest here because even though the numbers of nodes and links were halved, leaf links remained unchanged and became the shortest in the graph.
For less conservative time steps, GNM simulations take seconds to minutes versus hours to days for iPNM.

\subsection{GNM as a reduced-order model or theory}
\label{sec:disc_theory}
The ganglion network model can be viewed and used in two ways.
As a reduced-order model, which is how we presented it, GNM is a graph-based abstraction of all possible \textit{ganglion states} in the porous medium. In other words, any ganglion we can imagine fitting inside the void space is represented by a point on the graph. This graph is undirected and acyclic, or a tree,\footnote{Technically, the graph can consist of a collection of disconnected components (three in Fig.~\ref{fig:coarse}), called a \textit{forest}. However, since we place ganglia only on the largest connected component, wherein they can exchange mass through the intervening water, we have referred to the graph as a \textit{tree}.} governed by the rules of motion we postulated in Section~\ref{sec:rules} and validated in Section~\ref{sec:results}.
The pore-network model (PNM) is a similar abstraction, but its undirected graph does not possess any particular structure, the points on it represent all possible \textit{positions} the non-wetting phase can occupy, and it is governed by coupled balance equations written at the nodes.
Simulating with GNM is therefore equivalent to adopting a ganglion-centric viewpoint of the physics, compared to a pore-centric viewpoint offered by PNM. 

We can also view GNM as a theory, generalizing the LSW theory of Ostwald ripening~\cite{lifshitz1961orig, wagner1961orig} from spherical bubbles in a bulk fluid to trapped ganglia of arbitrary shape inside a porous medium, allowing also for dissolution and growth in addition to ripening.
LSW evolves the number density function of bubble radii living on a one-dimensional axis (i.e., positive real line), called the \textit{phase space}. Bueno et al.~\cite{bueno2024theory, bueno2025theory} showed that for populations whose bubbles occupy a single pore, the phase space remains Euclidean with $n_c\!+\!1$ axes, where $n_c$ is the number of chemical species comprising the bubble.
For general multi-pore occupying ganglia, the phase space remained unknown, because any Euclidean form demands an intractable number of axes that grows combinatorially with pore occupancy. This work showed that a tree graph serves as a parsimonious phase space for general ganglia trapped in a porous medium.

The theory framing is then completed by allowing an \textit{ensemble} of ganglia to evolve on this graph. Concretely, millions of ganglia can move on the \textit{same} graph extracted from a single porous sample. The only nuance here is that the ensemble consists of many sub-populations, each corresponding to a realization. Ganglia within a realization occupy the same copy of the sample in an infinite tiling of physical space like cells in a lattice. By carrying an identifying label, each sub-population's movement on the graph is governed by the rules of Section~\ref{sec:rules} (e.g., antichain).
All ganglia, even those from different realizations, interact via the mean field as outlined in Section~\ref{sec:meanfield}.
In this way, GNM can evolve ganglion populations at scales PNM cannot, and given GNM's agnosticism to the spatial scale, a path to upscaling emerges.
It is acknowledged that the common conception of a theory is one that has closed-form formulae, rather than an algorithmic recipe. Be that as it may, the logical extension of LSW leads to Algorithm~\ref{alg:gnm}.

\subsection{Limitations and extensions}
\label{sec:disc_limit}

The GNM formulated here has a few limitations that can be removed in future versions. We list them in order, each followed by a plausible avenue for extension.
First, the mean-field approximation discussed in Section~\ref{sec:disc_discrep} limits GNM to concentration fields whose volume average is spatially uniform (i.e., fluctuations around a mean are allowed, but the mean must have no trend).
If the mean does possess a trend, $X_m(\mathbf{x})$ where $\mathbf{x}$ is the position vector, GNM still applies and the volume-change rate of each ganglion can be computed from Eq.~\ref{eq:gangrate}.
Except now the mean field must be evaluated at the ganglion's spatial position, interpolatable from the nodal positions $\mathbf{x}_i$ of the tree graph. The $X_m(\mathbf{x})$ itself may be known as an input, or derived from a separate transport solver (e.g., PNM) coupled to GNM.
A second limitation is that our tree graphs were extracted using classical PMM, which assumes zero contact angle. For non-zero contact angles, recent generalizations of PMM can be used~\cite{schulz2015pmmCA, liu2022pmmCA} to build the graph, with all else in GNM kept intact. The real challenge is posed by mixed-wet domains, for which no obvious remedy is currently in sight.

Third, the capillary entry and snap-off curvatures stored at the virtual and junction nodes of the graph assume straight throats, which in 3D only have square cross sections. Improving on this approximation requires local image analysis, as part of the network extraction in Section~\ref{sec:extract}, on sub-images of throats (e.g., MS-P theory~\cite{mayer1965MSP, princen1969MSP}).
Incidentally, said limitation and remedy apply to PNM as well.
Fourth, we neglected the nucleation of new bubbles due to the supersaturation of water. In GNM, this can be modeled easily by spawning new ganglia at the leaf links, whose interaction and ripening with existing ganglia present an interesting line of inquiry.
Another limitation here is that solubility and diffusive mass transfer provided the driving force for ganglion evolution, whose dynamics on the tree graph obeyed the rules of Section~\ref{sec:rules}. Other forces---flow, gravity, centrifugal, and electromagnetic---can also induce ganglion mobilization within the pore space, but are governed by rules that differ fundamentally from those herein. For instance, if fast-flowing water mobilizes (but does not fragment) a ganglion from state A to state B, two points on different branches of the tree graph, then the volume of that ganglion must remain unchanged. But under the postulated rules, the only way for the ganglion to move between branches is for it to overshoot and spill at the common junction, which requires volume change. Hence, new rules are in need of discovery for such problems.

The above extensions could offer a fresh angle on studying whole families of problems involving the evolution of ganglia in porous media, such as depressurization-induced bubble exsolution~\cite{berg2020PressDeplete}, boiling~\cite{mori2009boil}, non-aqueous phase liquid (NAPL) dissolution~\cite{chomsurin2003NAPL}, and non-wetting droplet or bubble removal from fuel cells and electrolyzers~\cite{lee2020FCbazyl, lu2010FC}. They may also, with further extension, provide a new way of tackling geochemistry problems involving nucleation, dissolution, and precipitation of mineral crystals, which can exert crystallization stress on the porous solid~\cite{scherer1999crystal}.

\section{Conclusion}
\label{sec:conc}

We presented the ganglion network model (GNM), a new method for simulating how partially miscible ganglia trapped inside a porous medium evolve through diffusive mass transfer. GNM operates on a binary image of the medium (e.g., from X-ray micro-CT) and applies a modified pore-morphology method to derive a tree graph, the \textit{ganglion network}. Every point on this graph is a ganglion state with uniquely determined volume, curvature, and spatial position. Neither the microstructure nor the ganglia are geometrically simplified in the process of extracting the network.
In GNM, a population of ganglia is evolved as a set of points on the tree graph, subject to postulated rules of motion that capture capillary events such as invasion, snap-off, fragmentation, and merger.
The rate at which a ganglion moves on the graph is set by its mass exchange with a postulated mean field, which represents the water and is capable of storing dissolved solute. We compared GNM against a recent image-based pore-network model (iPNM)~\cite{laku2026ipnm}, itself validated against microfluidic experiments of Ostwald ripening of hydrogen bubbles~\cite{salehpour2025micro}. We considered 2.5D (planar with out-of-plane thickness) and 3D domains under three scenarios: ripening, dissolution, and growth. The two models agree well on the evolution of mean curvature, ganglion count, non-wetting-phase volume, and total moles, as well as on the number distributions of ganglion curvature and volume and the spatial configuration of ganglia over time. Similar to PNM, which is a graph-based, reduced-order abstraction of the pore space, GNM is a graph-based abstraction of ganglion states. In other words, every ganglion that fits within the pore space is represented by a point on the graph. Unlike PNM, whose cost scales nonlinearly with domain size and requires solving nonlinear systems, the cost of GNM scales linearly with the number of ganglia, is indifferent to domain size, and requires no solution of a system of equations. The limitations of the present formulation and its future extensions were discussed towards tackling a wide range of problems involving the evolution of trapped phases in porous media.

\section*{Acknowledgments}
This research is supported by the National Science Foundation, United States under Grant No. CBET-2348723.

\appendix
\setcounter{figure}{0}
\section{Entry curvature of the 2.5D virtual nodes}
\label{app:entry}

\begin{figure}[t!]
\centering
\includegraphics[width=0.55\linewidth]{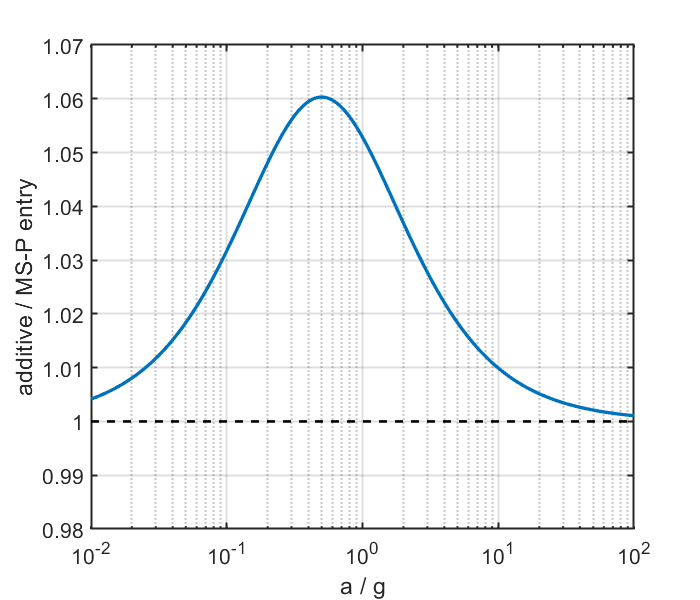}
\caption{Ratio of the additive virtual-node curvature $1/a + 2/g$ to the Mayer--Stowe--Princen entry curvature of a rectangular prism with cross section $a \times g$, at zero contact angle, as a function of the aspect ratio $a/g$. The dashed line marks perfect agreement.}
\label{fig:entrycomp}
\end{figure}

The virtual nodes of the 2.5D initial graph store the additive curvature $1/a + 2/g$ of Eq.~\ref{eq:nodecrv}, where $a$ is the in-plane throat radius and $g$ is the gap thickness. Section~\ref{sec:extract} leaves these values untouched on the grounds that they approximate the capillary entry curvature of the throat. Fig.~\ref{fig:entrycomp} substantiates this claim. It compares the additive value against the entry curvature of a prismatic throat with rectangular cross section, computed by the Mayer--Stowe--Princen method~\cite{mayer1965MSP,princen1969MSP} at zero contact angle, over four decades of the aspect ratio $a/g$. The ratio of the two remains within $[1.0, 1.06]$ throughout, so the additive value overestimates the true entry curvature by at most 6\%.

\section{Mass-conservative formulation of iPNM}
\label{app:ipnm}

iPNM~\cite{laku2026ipnm} writes three molar balance equations for each pore $i$. These include the wetting phase, the non-wetting component in the bubble phase, and the non-wetting component dissolved in the wetting phase:
\begin{align}
\rho_w V_i \frac{dS_{w,i}}{dt} &= \rho_w \sum_{j=1}^{z_i} q_{w,ij}, \label{eq:B_wet}\\
\rho_b V_i \frac{dS_{b,i}}{dt} &= \rho_b \sum_{j=1}^{z_i} q_{b,ij} + \dot{m}_i, \label{eq:B_bub}\\
\frac{d(\rho_w C_i S_{w,i} V_i)}{dt} &= \rho_w \sum_{j=1}^{z_i} J_{ij} - \dot{m}_i, \label{eq:B_dis}
\end{align}
where $V_i$ is the pore volume, $S_{w,i}$ and $S_{b,i}\!=\!1-S_{w,i}$ are the wetting and non-wetting saturations, $C_i$ is the mole fraction of dissolved non-wetting component, $\rho_w$ and $\rho_b$ are the molar densities of the phases, $q_{w,ij}$ and $q_{b,ij}$ are the volumetric flow rates through throat $ij$, $J_{ij}$ is the solute flux through the throat, and $\dot{m}_i$ is the interphase mass-transfer rate.

Eq.~\ref{eq:B_bub} is then operator split into a flow part (fast) and a mass-transfer part (slow). The flow part is:
\begin{equation}
\label{eq:B_bubflow}
\rho_b V_i \frac{dS_{b,i}}{dt} = \rho_b \sum_{j=1}^{z_i} q_{b,ij},
\end{equation}
and together with Eq.~\ref{eq:B_wet}, constitutes the \textit{flow problem}. Eqs.~\ref{eq:B_wet} and \ref{eq:B_bubflow} are solved for $P_{w,i}$ and $S_{b,i}$ while keeping $C_i$ frozen. The mass-transfer part, $\rho_b V_i\, dS_{b,i}/dt\!=\!\dot{m}_i$, is added to Eq.~\ref{eq:B_dis} to eliminate $\dot{m}_i$. This yields:
\begin{equation}
\label{eq:B_total}
\rho_b V_i \frac{dS_{b,i}}{dt} + \frac{d(\rho_w C_i S_{w,i} V_i)}{dt} = \rho_w \sum_{j=1}^{z_i} J_{ij},
\end{equation}
which constitutes the \textit{transport problem}, and it is solved for $C_i$ and $S_{b,i}$ while keeping $q_{w,ij}$ frozen.

In~\cite{laku2026ipnm}, Eq.~\ref{eq:B_total} is simplified for bubble-occupied pores before solving it. Namely, the second term on the left-hand side (LHS) is dropped (Eq.~10 therein) on order-of-magnitude grounds ($\rho_w C_i \ll \rho_b$). However, as remarked in~\cite{laku2026ipnm}, the omission makes the scheme non-conservative, since the dissolved solute swept by a moving interface is lost. Throughout this work, we have retained this term, rendering iPNM conservative to machine precision.

For a vacant pore, the first term on the LHS of Eq.~\ref{eq:B_total} is zero and the equation reduces to the standard transport of $C_i$.
For an occupied pore, the concentration is set by local equilibrium, $C_i\!=\!C_{eq}(S_{b,i})$, through the capillary pressure of the pore and Henry's law (Section~3.4 of~\cite{laku2026ipnm}). Thus, the full set of unknowns is the mixed vector:
\begin{equation}
\label{eq:B_unk}
x_i = \begin{cases} S_{b,i}, & \text{occupied pore},\\ C_i, & \text{vacant pore}.\end{cases}
\end{equation}

Discretizing the time derivative in Eq.~\ref{eq:B_total} via backward Euler over the sub-step $\delta t$ yields the residual:
\begin{equation}
\label{eq:B_res}
R_i = \begin{cases}
\tfrac{\rho_b}{\rho_w}\big(S_{b,i}^{n+1} - S_{b,i}^{n}\big) + \big(C_i S_{w,i}\big)^{n+1} - \big(C_i S_{w,i}\big)^{n} - \dfrac{\delta t}{V_i}\displaystyle\sum_{j=1}^{z_i} J_{ij}^{n+1}, & \text{occupied pore},\\[10pt]
C_i^{n+1} - C_i^{n} - \dfrac{\delta t}{V_i}\displaystyle\sum_{j=1}^{z_i} J_{ij}^{n+1}, & \text{vacant pore},
\end{cases}
\end{equation}
for pore $i$. The corresponding nonlinear system, obtained by writing Eq.~\ref{eq:B_res} for all pores, is solved using Newton's method using automatic differentiation. Solute is transported with flux $J_{ij}$ through every throat, including occupied ones via corner films. Moreover, $J_{ij}$ consists of contributions from advection (whose $q_{w,ij}$ is frozen) and diffusion.

After every accepted sub-step, the capillary stability check of~\cite{laku2026ipnm} (invasion, snap-off, and retraction) is performed, followed by a \textit{flash} calculation at every pore affected by an event. The flash restores local equilibrium while conserving total non-wetting moles $N_i$ in the pore. It consists of the following root-finding problem for $S_{b,i}$:
\begin{equation}
\label{eq:B_flash}
\tfrac{\rho_b}{\rho_w}\, S_{b,i} + C_{eq}(S_{b,i})\,(1 - S_{b,i}) = \frac{N_i}{\rho_w V_i}
\end{equation}
where the LHS is the sum of the moles of the non-wetting phase and the dissolved moles in water, both functions of $S_{b,i}$. The right-hand side (RHS) is the (normalized) number of total moles in the pore, which is known. The root is found via bisection.
If no root exists, the non-wetting phase is too small and dissolves into the water in the pore. The flash is needed for robust convergence, since small bubbles have high capillary pressure that can derail Newton.
Lastly, Newton updates are capped so that $S_{b,i}$ does not drop below a small threshold, and a $\delta t$ that fails to converge is rejected and halved.
The closed-boundary ripening runs of Section~\ref{sec:res_ripen} show $N_t$ is constant, confirming conservation.

\biboptions{numbers,sort&compress}
\bibliographystyle{elsarticle-num}
\bibliography{./References}

\end{document}